\RequirePackage{fix-cm}
\documentclass[smallextended]{svjour3}       % onecolumn (second format)
\smartqed  % flush right qed marks, e.g. at end of proof
\usepackage{graphicx}
\journalname{Empirical Software Engineering}
\usepackage[authoryear]{natbib}
\usepackage{multirow}
\usepackage{array}
\usepackage{enumitem}
\newcolumntype{C}[1]{>{\centering\arraybackslash}p{#1}}
\usepackage{colortbl}
\usepackage{hyperref}% http://ctan.org/pkg/hyperref

\usepackage{float} % force figure location
\usepackage{graphicx}
\usepackage{adjustbox}
\usepackage[font=small,labelfont=bf,tableposition=top]{caption}
\usepackage{newfloat}
\usepackage{balance}
\DeclareCaptionType{Stack Overflow post}
\usepackage[T1]{fontenc} 
\usepackage[utf8]{inputenc}
\usepackage{subfig}
\usepackage[tikz]{bclogo}
\usepackage{diagbox}
\usepackage{listings}
\usepackage{xargs}    
\usepackage{xspace}
\usepackage{wrapfig}
\usepackage{tcolorbox} % text box
\usepackage{hhline}
\usepackage{tikz}
\usepackage{scalerel}

\newcommand{\tikztriangleright}[1][red,fill=red]{\scalerel*{\tikz \draw[rounded corners=0.1pt,#1] (0,-2.5pt)--++(0,5pt)--++(-30:5pt)--cycle;}{\triangleright}}

\newcommand{\tikzcircle}[2][red,fill=red]{\tikz[baseline=-0.5ex]\draw[#1,radius=#2] (0,0) circle ;}%

\usepackage{amsmath,amssymb,amsfonts}
\AtBeginDocument{%
	\providecommand\BibTeX{{%
			\normalfont B\kern-0.5em{\scshape i\kern-0.25em b}\kern-0.8em\TeX}}}

\begin{document}
% generally useful macros for writing
\newcommand{\TexDir}{/home3/aiken/tex}

% [[[There should really be a list environment for defining research questions, so that they can be given labels.]]]
\newcommand{\RQ}[1]{\textbf{RQ#1}\xspace}

% EMSE revision
\newcommand{\revision}[1]{\textcolor{blue}{#1}}

% These allow switching interline spacing; the change takes effect immediately:

\makeatletter
\newcommand{\singlespacing}{\let\CS=\@currsize\renewcommand{\baselinestretch}{1}\tiny\CS}
\newcommand{\oneandahalfspacing}{\let\CS=\@currsize\renewcommand{\baselinestretch}{1.25}\tiny\CS}
\newcommand{\doublespacing}{\let\CS=\@currsize\renewcommand{\baselinestretch}{1.5}\tiny\CS}
%setspacingto sets the interline spacing to the value of its argument
% e.g., \setspacingto{1.5} is the same as \doublespacing
\newcommand{\setspacingto}[1]{\let\CS=\@currsize\renewcommand{\baselinestretch}{#1}\tiny\CS}
\makeatother

% nonumber
\newcommand{\nn}{\nonumber}

% Tab for hand-formatting:
\newcommand{\tab}{\hspace*{2em}}

% Angle brackets:
\newcommand{\la}{\langle}
\newcommand{\ra}{\rangle}

% s.t.
\newcommand{\st}{\mbox{\ s.t.\ }}

% otherwise
\newcommand{\ow}{\m{\rm otherwise}}

% if
\newcommand{\mif}{\m{\rm if\ }}

% Harpoons
\newcommand{\rh}{\rightharpoonup}
\newcommand{\lh}{\leftharpoonup}

% Denotational-semantics-style brackets and bottom:
\newcommand{\lbk}{\lbrack\!\lbrack}
\newcommand{\rbk}{\rbrack\!\rbrack}
\newcommand{\bottom}{\perp}

% Projection operator
\newcommand{\proj}{\!\downarrow\!}

% macros for mbox combined with another style
% (useful for changing typefaces in math mode)
\newcommand {\mboxbf}[1]{\mbox{{\bf #1}}}
\newcommand {\mboxit}[1]{\mbox{{\it #1}}}
\newcommand {\mboxem}[1]{\mbox{{\em #1}}}
\newcommand {\m}{\mbox}
\newcommand {\ch}{\rm}
\newcommand{\change}[1]{{\color{red!60!white}\textbf{}~#1}\xspace}
	
% macro for creating a binary operator
%
% example:  \makebinop{\makebinop{\mybmod}{mod}
%	(duplicates the \bmod macro)
%
\def\makebinop#1#2{\def#1{\mskip-\medmuskip \mskip5mu
\mathbin{\rm #2} \penalty900 \mskip5mu \mskip-\medmuskip}}

% Environments for theorems, lemmas, etc.
%\newtheorem{theorem}{Theorem}[section]
%\newtheorem{lemma}[theorem]{Lemma}
%\newtheorem{corollary}[theorem]{Corollary}
%\newtheorem{definition}[theorem]{Definition}

%\newtheorem{observation}[theorem]{Observation}
%\newtheorem{fact}{Fact}[theorem]

%\newtheorem{proposition}[theorem]{Proposition}
%\newtheorem{example}[theorem]{Example}

%\newtheorem{constraint}[theorem]{Constraint}
%\newtheorem{axiom}[theorem]{Axiom}
%\newtheorem{law}[theorem]{Law}
%\newtheorem{algorithm}[theorem]{Algorithm}
%\newtheorem{invariant}[theorem]{Invariant}

% Definition of the proof-environment:
%\newenvironment{proof}{{\bf Proof:}\quad}{$\Box$}

% Definition of an eqnarray-like environment with an extra
% column for comments
\newcommand{\eeqnarray}[1]{\[\begin{array}{rcll} #1 \end{array}\]}

% Definition of a eqnarray-like environment for proofs of the
% form 
%     init     
%=>   step1   reason1
%=>   step2   reason2
%...
\newcommand{\peqnarray}[1]{\[\begin{array}{cll} #1 \end{array}\]}

%
% An environment for formatting programs. 
%
%	written by Hal Perkins
%	adapted by Charles Elkan	9/3/86
%	keyword macros by Anne Neirynck
%
% Usage :
%
% \begin{program}
% program text\\
% program text
% \end{program}
%
% The program environment is a tabbing environment with ten tab stops spaced
% evenly from the left of the page.  Initially the left margin is the second
% tab stop.  Use \+ to indent following lines one more tab stop, \- to undo
% the effect of \+, and \> at the beginning of a line to indent an extra tab.

\newlength{\pgmtab}          %  \pgmtab is the width of each tab in the
\setlength{\pgmtab}{2em}     %  program environment

% boxed program is like program, only boxed!
% This is useful for centering programs and preventing page breaks in programs.
% argument t or b is required.

\newenvironment{boxed-program}[1]{\begin{minipage}[#1]{9in}
\begin{tabbing}\hspace{\pgmtab}\=\hspace{\pgmtab}\=%
\hspace{\pgmtab}\=\hspace{\pgmtab}\=\hspace{\pgmtab}\=\hspace{\pgmtab}\=%
\hspace{\pgmtab}\=\hspace{\pgmtab}\=\hspace{\pgmtab}\=\hspace{\pgmtab}\=%
\hspace{\pgmtab}\=%
\kill}{\end{tabbing}\end{minipage}}

\newenvironment{program}{\begin{tabbing}\hspace{\pgmtab}\=\hspace{\pgmtab}\=%
\hspace{\pgmtab}\=\hspace{\pgmtab}\=\hspace{\pgmtab}\=\hspace{\pgmtab}\=%
\hspace{\pgmtab}\=\hspace{\pgmtab}\=\hspace{\pgmtab}\=\hspace{\pgmtab}\=%
\hspace{\pgmtab}\=%
\+\+\kill}{\end{tabbing}}

% The following commands should be used OUTSIDE math mode

\newcommand {\FUNCTION}{{\bf function\ }}
\newcommand {\BEGIN}{{\bf begin\ }}
\newcommand {\END}{{\bf end}}
\newcommand {\CASE}{{\bf case}}
\newcommand {\OF}{{\bf of}}
\newcommand {\SELECT}{{\bf select\ }}
\newcommand {\WHERE}{{\bf where\ }}
\newcommand {\DECLARE}{{\bf declare\ }}
\newcommand {\ARRAY}{{\bf array\ }}
\newcommand {\LET}{{\bf\ let\ }}
\newcommand {\IN}{{\bf\ in\ }}
\newcommand {\IF}{{\bf if\ }}
\newcommand {\THEN}{{\bf then\ }}
\newcommand {\ELSE}{{\bf else\ }}
\newcommand {\SKIP}{{\bf skip\ }}
\newcommand {\DO}{{\bf do\ }}
% OLD---don't use \OD
\newcommand {\OD}{{\bf od\ }}
\newcommand {\BY}{{\bf by\ }}
\newcommand {\LOOP}{{\bf loop\ }}
\newcommand {\WHILE}{{\bf while\ }}
\newcommand {\TO}{{\bf to\ }}
\newcommand {\DOWNTO}{{\bf down to\ }}
\newcommand {\FOR}{{\bf for\ }}
\newcommand {\FOREACH}{{\bf for each\ }}
\newcommand {\RETURN}{{\bf return\ }}
\newcommand {\REPEAT}{{\bf repeat\ }}
\newcommand {\UNTIL}{{\bf until\ }}
\newcommand {\LCOM}{$(\ast \;$}
\newcommand {\RCOM}{$\ast)$}
\newcommand {\GOTO}{{\bf goto\ }}

% The following commands are for use INSIDE math mode

\newcommand {\OP}[1]{\mbox{\sc #1}}
\newcommand {\op}[1]{\mbox{\sc #1}}
\newcommand {\mm}[1]{\mbox{\rm #1}\;}
\newcommand {\id}[1]{\mbox{\it #1\ }}

\newcommand {\ASSIGN}{\leftarrow }
\newcommand {\MIN}{\OP{min} }
\newcommand {\MOD}{\; {\bf{\rm mod}} \; } 
\newcommand {\LAMBDA}{{\bf \lambda\ }}
\newcommand {\FALSE}{{\em FALSE\ }}

% Miscellaneous notation

%\newcommand {\And}{\wedge}
\newcommand {\Or}{\vee}

\newcommand {\thus}{{\dot{. \: .}\;}}

\newcommand {\bigO}[1]{{\cal O}(#1)}

\newcommand {\app}{\!\!:\!}
\newcommand {\hastype}{::}
\newcommand{\hast}{:}
\newcommand{\qt}[1]{\mbox{``#1''}}
\newcommand{\TexComment}[1]{}

\newcommand{\dq}{\m{\tt "}}
\newcommand{\flatqt}[1]{\m{\dq #1 \dq}}
\newcommand{\seq}{\subseteq}
\newcommand{\derives}{\vdash}

% for writing grammars
\newcommand{\grammar}{::=}
\newcommand{\gor}{\,|\,}

% macros for writing inference rules
\newcommand{\infrule}[2]{\displaystyle{\displaystyle\strut{#1}} \over %
                                        {\displaystyle\strut {#2}}}
\newcommand{\cinfrule}[3]{\parbox{14cm}{\hfil$\infrule{#1}{#2}$\hfil}\parbox{4cm}{$\,#3$\hfil}}

% ASE'20 paper related
\newcommand{\SO}{\emph{SO}\xspace}
% [[[Should ML really be in italics? -- Gary]]]
\newcommand{\ML}{ML\xspace}
\newcommand{\tf}{\emph{TensorFlow}\xspace}
\newcommand{\scikit}{\emph{Scikit-learn}\xspace}
\newcommand{\keras}{\emph{Keras}\xspace}
\newcommand{\torch}{\emph{PyTorch}\xspace}
\newcommand{\etal}{\emph{et al.}\xspace}
\newcommandx{\improve}[2][1=]{\todo[linecolor=red,backgroundcolor=red!25,bordercolor=red,#1]{#2}}

%color
\definecolor{Gray}{gray}{0.85}
\definecolor{deepblue}{rgb}{0,0,0.5}
\definecolor{deepred}{rgb}{0.6,0,0}
\definecolor{deepgreen}{rgb}{0,0.5,0}
\definecolor{lightorange}{rgb}{1, 0.7804, 0.0078}
\definecolor{lightgreen}{rgb}{.1961,.7961,0}
\definecolor{lightblue}{rgb}{.4078,.7961,.8157}

% listing ML codes
\newsavebox{\mybox}
\definecolor{halfgray}{gray}{0.55}
\definecolor{ipython_frame}{RGB}{207, 207, 207}
\definecolor{ipython_bg}{RGB}{247, 247, 247}
\definecolor{ipython_red}{RGB}{186, 33, 33}
\definecolor{ipython_green}{RGB}{0, 128, 0}
\definecolor{ipython_cyan}{RGB}{64, 128, 128}
\definecolor{ipython_purple}{RGB}{170, 34, 255}
\lstset{
	breaklines=true,
	extendedchars=true,
	literate=
	{á}{{\'a}}1 {é}{{\'e}}1 {í}{{\'i}}1 {ó}{{\'o}}1 {ú}{{\'u}}1
	{Á}{{\'A}}1 {É}{{\'E}}1 {Í}{{\'I}}1 {Ó}{{\'O}}1 {Ú}{{\'U}}1
	{à}{{\`a}}1 {è}{{\`e}}1 {ì}{{\`i}}1 {ò}{{\`o}}1 {ù}{{\`u}}1
	{À}{{\`A}}1 {È}{{\'E}}1 {Ì}{{\`I}}1 {Ò}{{\`O}}1 {Ù}{{\`U}}1
	{ä}{{\"a}}1 {ë}{{\"e}}1 {ï}{{\"i}}1 {ö}{{\"o}}1 {ü}{{\"u}}1
	{Ä}{{\"A}}1 {Ë}{{\"E}}1 {Ï}{{\"I}}1 {Ö}{{\"O}}1 {Ü}{{\"U}}1
	{â}{{\^a}}1 {ê}{{\^e}}1 {î}{{\^i}}1 {ô}{{\^o}}1 {û}{{\^u}}1
	{Â}{{\^A}}1 {Ê}{{\^E}}1 {Î}{{\^I}}1 {Ô}{{\^O}}1 {Û}{{\^U}}1
	{œ}{{\oe}}1 {Œ}{{\OE}}1 {æ}{{\ae}}1 {Æ}{{\AE}}1 {ß}{{\ss}}1
	{ç}{{\c c}}1 {Ç}{{\c C}}1 {ø}{{\o}}1 {å}{{\r a}}1 {Å}{{\r A}}1
	{€}{{\EUR}}1 {£}{{\pounds}}1
}

%%
%% Python definition (c) 1998 Michael Weber
%% Additional definitions (2013) Alexis Dimitriadis
%% modified by me (should not have empty lines)
%%
\lstdefinelanguage{iPython}{
	morekeywords={access,and,break,class,continue,def,del,elif,else,except,exec,finally,for,from,global,if,import,in,is,lambda,not,or,pass,print,raise,return,try,while,KerasRegressor,DataLoader,Dense},%
	%
	% Built-ins
	morekeywords=[2]{abs,all,any,basestring,bin,bool,bytearray,callable,chr,classmethod,cmp,compile,complex,delattr,dict,dir,divmod,enumerate,eval,execfile,file,filter,float,format,frozenset,getattr,globals,hasattr,hash,help,hex,id,input,int,isinstance,issubclass,iter,len,list,locals,long,map,max,memoryview,min,next,object,oct,open,ord,pow,property,range,raw_input,reduce,reload,repr,reversed,round,set,setattr,slice,sorted,staticmethod,str,sum,super,tuple,type,unichr,unicode,vars,xrange,zip,apply,buffer,coerce,intern},%
	sensitive=true,%
	morecomment=[l]\#,%
	morestring=[b]',%
	morestring=[b]",%
	morestring=[s]{'''}{'''},% used for documentation text (mulitiline strings)
	morestring=[s]{"""}{"""},% added by Philipp Matthias Hahn
	morestring=[s]{r'}{'},% `raw' strings
	morestring=[s]{r"}{"},%
	morestring=[s]{r'''}{'''},%
	morestring=[s]{r"""}{"""},%
	morestring=[s]{u'}{'},% unicode strings
	morestring=[s]{u"}{"},%
	morestring=[s]{u'''}{'''},%
	morestring=[s]{u"""}{"""},%
	%
	% {replace}{replacement}{lenght of replace}
	% *{-}{-}{1} will not replace in comments and so on
	literate=
	*{+}{{{\color{ipython_purple}+}}}1
	{-}{{{\color{ipython_purple}-}}}1
	{*}{{{\color{ipython_purple}$^\ast$}}}1
	{/}{{{\color{ipython_purple}/}}}1
	{^}{{{\color{ipython_purple}\^{}}}}1
	{?}{{{\color{ipython_purple}?}}}1
	{!}{{{\color{ipython_purple}!}}}1
	{\%}{{{\color{ipython_purple}\%}}}1
	{<}{{{\color{ipython_purple}<}}}1
	{>}{{{\color{ipython_purple}>}}}1
	{|}{{{\color{ipython_purple}|}}}1
	{\&}{{{\color{ipython_purple}\&}}}1
	{~}{{{\color{ipython_purple}~}}}1
	{==}{{{\color{ipython_purple}==}}}2
	{<=}{{{\color{ipython_purple}<=}}}2
	{>=}{{{\color{ipython_purple}>=}}}2
	{+=}{{{+=}}}2
	{-=}{{{-=}}}2
	{*=}{{{$^\ast$=}}}2
	{/=}{{{/=}}}2,
	literate=
	{á}{{\'a}}1 {é}{{\'e}}1 {í}{{\'i}}1 {ó}{{\'o}}1 {ú}{{\'u}}1
	{Á}{{\'A}}1 {É}{{\'E}}1 {Í}{{\'I}}1 {Ó}{{\'O}}1 {Ú}{{\'U}}1
	{à}{{\`a}}1 {è}{{\`e}}1 {ì}{{\`i}}1 {ò}{{\`o}}1 {ù}{{\`u}}1
	{À}{{\`A}}1 {È}{{\'E}}1 {Ì}{{\`I}}1 {Ò}{{\`O}}1 {Ù}{{\`U}}1
	{ä}{{\"a}}1 {ë}{{\"e}}1 {ï}{{\"i}}1 {ö}{{\"o}}1 {ü}{{\"u}}1
	{Ä}{{\"A}}1 {Ë}{{\"E}}1 {Ï}{{\"I}}1 {Ö}{{\"O}}1 {Ü}{{\"U}}1
	{â}{{\^a}}1 {ê}{{\^e}}1 {î}{{\^i}}1 {ô}{{\^o}}1 {û}{{\^u}}1
	{Â}{{\^A}}1 {Ê}{{\^E}}1 {Î}{{\^I}}1 {Ô}{{\^O}}1 {Û}{{\^U}}1
	{œ}{{\oe}}1 {Œ}{{\OE}}1 {æ}{{\ae}}1 {Æ}{{\AE}}1 {ß}{{\ss}}1
	{ç}{{\c c}}1 {Ç}{{\c C}}1 {ø}{{\o}}1 {å}{{\r a}}1 {Å}{{\r A}}1
	{€}{{\EUR}}1 {£}{{\pounds}}1,
	%
	%   identifierstyle=\color{red}\ttfamily,
	commentstyle=\color{ipython_cyan}\ttfamily,
	stringstyle=\color{ipython_red}\ttfamily,
	keepspaces=true,
	showspaces=false,
	showstringspaces=false,
	rulecolor=\color{black},
	frame=single,
	frameround={t}{t}{t}{t},
	framexleftmargin=3.2mm,
	numbers=left,
	numberstyle=\tiny\color{halfgray},
	backgroundcolor=\color{ipython_bg},
	%   extendedchars=true,
	basicstyle=\scriptsize\ttfamily,
	keywordstyle=\color{ipython_green}\ttfamily,
}

%% print error message
\lstdefinelanguage{iPython2}{
	morekeywords={access,and,break,class,continue,def,del,elif,else,except,exec,finally,for,from,global,if,import,in,is,lambda,not,or,pass,print,raise,return,try,while,IndexError,Traceback,Sequential},%
	%
	% Built-ins
	morekeywords=[2]{abs,all,any,basestring,bin,bool,bytearray,callable,chr,classmethod,cmp,compile,complex,delattr,dict,dir,divmod,enumerate,eval,execfile,file,filter,float,format,frozenset,getattr,globals,hasattr,hash,help,hex,id,int,isinstance,issubclass,iter,len,locals,long,map,max,memoryview,min,next,object,oct,open,ord,pow,property,raw_input,reduce,reload,repr,reversed,round,set,setattr,slice,sorted,staticmethod,str,sum,super,tuple,type,unichr,unicode,vars,xrange,zip,apply,buffer,coerce,intern},%
	sensitive=true,%
	morecomment=[l]\#,%
	morestring=[b]',%
	morestring=[b]",%
	morestring=[s]{'''}{'''},% used for documentation text (mulitiline strings)
	morestring=[s]{"""}{"""},% added by Philipp Matthias Hahn
	morestring=[s]{r'}{'},% `raw' strings
	morestring=[s]{r"}{"},%
	morestring=[s]{r'''}{'''},%
	morestring=[s]{r"""}{"""},%
	morestring=[s]{u'}{'},% unicode strings
	morestring=[s]{u"}{"},%
	morestring=[s]{u'''}{'''},%
	morestring=[s]{u"""}{"""},%
	%
	% {replace}{replacement}{lenght of replace}
	% *{-}{-}{1} will not replace in comments and so on
	literate=
	*{+}{{{\color{ipython_purple}+}}}1
	{-}{{{\color{ipython_purple}-}}}1
	{*}{{{\color{ipython_purple}$^\ast$}}}1
	{/}{{{\color{ipython_purple}/}}}1
	{^}{{{\color{ipython_purple}\^{}}}}1
	{?}{{{\color{ipython_purple}?}}}1
	{!}{{{\color{ipython_purple}!}}}1
	{\%}{{{\color{ipython_purple}\%}}}1
	{<}{{{\color{ipython_purple}<}}}1
	{>}{{{\color{ipython_purple}>}}}1
	{|}{{{\color{ipython_purple}|}}}1
	{\&}{{{\color{ipython_purple}\&}}}1
	{~}{{{\color{ipython_purple}~}}}1
	{==}{{{\color{ipython_purple}==}}}2
	{<=}{{{\color{ipython_purple}<=}}}2
	{>=}{{{\color{ipython_purple}>=}}}2
	{+=}{{{+=}}}2
	{-=}{{{-=}}}2
	{*=}{{{$^\ast$=}}}2
	{/=}{{{/=}}}2,
	literate=
	{á}{{\'a}}1 {é}{{\'e}}1 {í}{{\'i}}1 {ó}{{\'o}}1 {ú}{{\'u}}1
	{Á}{{\'A}}1 {É}{{\'E}}1 {Í}{{\'I}}1 {Ó}{{\'O}}1 {Ú}{{\'U}}1
	{à}{{\`a}}1 {è}{{\`e}}1 {ì}{{\`i}}1 {ò}{{\`o}}1 {ù}{{\`u}}1
	{À}{{\`A}}1 {È}{{\'E}}1 {Ì}{{\`I}}1 {Ò}{{\`O}}1 {Ù}{{\`U}}1
	{ä}{{\"a}}1 {ë}{{\"e}}1 {ï}{{\"i}}1 {ö}{{\"o}}1 {ü}{{\"u}}1
	{Ä}{{\"A}}1 {Ë}{{\"E}}1 {Ï}{{\"I}}1 {Ö}{{\"O}}1 {Ü}{{\"U}}1
	{â}{{\^a}}1 {ê}{{\^e}}1 {î}{{\^i}}1 {ô}{{\^o}}1 {û}{{\^u}}1
	{Â}{{\^A}}1 {Ê}{{\^E}}1 {Î}{{\^I}}1 {Ô}{{\^O}}1 {Û}{{\^U}}1
	{œ}{{\oe}}1 {Œ}{{\OE}}1 {æ}{{\ae}}1 {Æ}{{\AE}}1 {ß}{{\ss}}1
	{ç}{{\c c}}1 {Ç}{{\c C}}1 {ø}{{\o}}1 {å}{{\r a}}1 {Å}{{\r A}}1
	{€}{{\EUR}}1 {£}{{\pounds}}1,
	%
	%   identifierstyle=\color{red}\ttfamily,
	commentstyle=\color{ipython_cyan}\ttfamily,
	stringstyle=\color{ipython_red}\ttfamily,
	keepspaces=true,
	showspaces=false,
	showstringspaces=false,
	rulecolor=\color{black},
	frame=single,
	frameround={t}{t}{t}{t},
	framexleftmargin=1.2mm,
	framexrightmargin=2.2mm,
	numbers=none,
	%numberstyle=\tiny\color{halfgray},
	%
	%
	backgroundcolor=\color{ipython_bg},
	%   extendedchars=true,
	basicstyle=\scriptsize\ttfamily,
	keywordstyle=\color{ipython_green}\ttfamily,
}
% end listing ML codes

%edit tcolorbox
% new tcolorbox environment
% #1: tcolorbox options
% #2: color
% #3: box title
\newtcolorbox{mytbox}[3][]
{
	colframe = #2!25,
	colback  = #2!12,
	coltitle = #2!8!black,  
	title    = {#3},
	#1,
}

\newtcolorbox{mytbox2}[2][]
{
	colframe = green!5!white,
	colback  = green!75!black,
	fonttitle=\scriptsize\bfseries,
	colbacktitle = green!85!black,
	title    = {#2},
	#1,
}

\newcolumntype{g}{>{\columncolor{Gray}}l}

\definecolor{codegreen}{rgb}{0,0.6,0}
\definecolor{codegray}{rgb}{0.5,0.5,0.5}
\definecolor{codepurple}{rgb}{0.58,0,0.82}
\definecolor{backcolour}{rgb}{0.95,0.95,0.92}

\def\checkmark{\tikz\fill[scale=0.4](0,.35) -- (.25,0) -- (1,.7) -- (.25,.15) -- cycle;} 

%\lstdefinestyle{mystyle}{
%	backgroundcolor=\color{backcolour},   
%	commentstyle=\color{codegreen},
%	keywordstyle=\color{magenta},
%	numberstyle=\tiny\color{codegray},
%	stringstyle=\color{codepurple},
%	basicstyle=\scriptsize,
%	breakatwhitespace=false,         
%	breaklines=true,                 
%	captionpos=b,                    
%	keepspaces=true,                 
%	numbers=left,                    
%	numbersep=5pt,                  
%	showspaces=false,                
%	showstringspaces=false,
%	showtabs=false,                  
%	tabsize=2
%}
%\lstset{style=mystyle}

\newcounter{NumObservations}
\stepcounter{NumObservations}
\newcommand{\finding}[1]{
	\begin{bclogo}[couleur= orange!5, epBord= 0.8, arrondi=0.2, logo=\bccrayon,marge= 1.2, ombre=true, blur, couleurBord=brown!90, tailleOndu=3, sousTitre ={\em #1}]{~Finding \arabic{NumObservations} $:$ } 
		
	\end{bclogo}
	\stepcounter{NumObservations}
}

\setlength{\fboxsep}{1pt}

\title{What Kinds of Contracts Do ML APIs Need? 
}


\author{Samantha Syeda Khairunnesa  \and 
	Shibbir Ahmed  \and 
	Sayem Mohammad Imtiaz \and 
	Hridesh Rajan \and 
	Gary T. Leavens
}


\institute{Samantha Syeda Khairunnesa  \at
              Dept. of Computer Science and Information Systems, Bradley University \\
              \email{skhairunnesa@fsmail.bradley.edu}           %  \\
           \and
           Shibbir Ahmed  \at
           Dept. of Computer Science, Iowa State University \\
           \email{shibbir@iastate.edu} 
           \and
           Sayem Mohammad Imtiaz  \at
           Dept. of Computer Science, Iowa State University \\
           \email{sayem@iastate.edu} 
           \and
           Hridesh Rajan  \at
           Dept. of Computer Science, Iowa State University \\
           \email{hridesh@iastate.edu} 
           \and
           Gary T. Leavens  \at
           Dept. of Computer Science, University of Central Florida \\
           \email{Leavens@ucf.edu} 
}

\date{Received: date / Accepted: date}

\maketitle

\begin{abstract}
%\input{abstract}	
%\begin{abstract}
Recent work has shown that Machine Learning (\ML)~programs are error-prone and called for contracts for {\ML} code. Contracts, as in the design by contract methodology, help document APIs and aid API users in writing correct code. 

The question is: what kinds of contracts would provide the most help to API users? 
We are especially interested in what kinds of contracts help API users
catch errors at earlier stages in the \ML~pipeline. 
We describe an empirical study of posts on 
\emph{Stack Overflow}
of the four most often-discussed \ML~libraries: 
\tf, \scikit, \keras, and \torch. 
For these libraries, our study extracted 413 informal (English) API specifications.
We used these specifications to understand the following questions. 
What are the root causes and effects behind \ML contract violations? 
Are there common patterns of \ML contract violations? 

When does understanding \ML contracts require an advanced level of \ML software expertise?
Could checking contracts at the API level help detect the violations in early \ML~pipeline stages? 
Our key findings are that the most commonly needed contracts for \ML~APIs are either checking constraints on single arguments of an API or
on the order of API calls.
The software engineering community could employ existing contract mining approaches to mine 
these contracts to promote an increased understanding of \ML~APIs.
We also noted a need to combine
behavioral and temporal contract mining approaches.
We report on categories of required \ML~contracts, which may help designers of contract languages.

\end{abstract}

\section{Introduction}
\label{sec:intro}

Software developers are increasingly integrating machine learning (\ML) 
into systems using ML libraries' application programming interfaces (APIs).
However, ML software is bug-prone ~\cite{10.1145/3213846.3213866, islam19, dfaults} and
like traditional software could benefit from adopting a design-by-contract methodology~\cite{islam19}.
Contracts can specify the expected behavior of an API and help client code use the API correctly,
e.g., a contract might require that the \texttt{fit} method be applied to a model before calling the \texttt{predict} method. Another example can be given using the \texttt{MaxPooling2D} method for retaining the most prominent features of the feature map in a convolutional neural network (CNN). There is a contract on the \texttt{MaxPooling2D} method's argument, \texttt{data\_format}, based on the shape of the input image. If the input image has the shape (N, C, H, W), then the value for the argument \texttt{data\_format} is set to \texttt{channels\_first}. If the input has the shape (N, H, W, C), then \texttt{data\_format} must be set to \texttt{channels\_last}. Here, the letters N, H, W, and C represent the following: the number of images in the batch, the height of the image, the width of the image, and the number of channels of the image.

There is a rich body of prior work 
that can be 
grouped into two categories: work on contracts for non-\ML software and work on \ML software.

The first category, contracts for non-\ML software,
can be further divided into two types: behavioral and temporal.
Behavioral contracts \cite{10.1145/363235.363259, Meyer88, 10.1109/ASE.2009.60, 10.1145/1858996.1859035, nguyen2014mining, khairunnesa2017exploiting}
specify acceptable program states, typically for calls to individual methods in an API. For instance, in the Java Development Kit (JDK) String class, the precondition `\texttt{beginIndex}$<=$\texttt{endIndex}' must be true before calling method \texttt{subString(beginIndex,endIndex)}. The contracts that belong to this category are preconditions (as in the example), or postconditions (constraints ensured by the execution of the call) for a method in question. There are also class invariants that capture the constraints for all methods in a particular class.
Temporal contracts \cite{Manna-Pnueli92, 10.1145/1831708.1831723, 10.1145/1595696.1595767, 10.1145/1287624.1287632} 
encode the correct ordering of calls, possibly among multiple APIs. For example, in Python, after creating a \texttt{threading.Lock} object, once a thread makes a call to \texttt{Lock.acquire()}, that thread should  eventually call \texttt{Lock.release()}. 

The notion of contracts in this study is similar to the kinds of contracts described just before this phrase. We have used the same definition of (behavioral and temporal) contracts in this study.
A contract specifies the correct usage of an API and an incorrect usage is a contract violation.

The second category is about \ML~software and its bugs \cite{10.1145/3213846.3213866, islam19, dfaults} 
and bug fixes \cite{8305957, islam20repairing}. 
These works study either the implementation of \ML library APIs or usage information about those APIs. \cite{10.1145/3213846.3213866} and \cite{dfaults} focused on understanding the defects in different ML libraries. The authors (\cite{10.1145/3213846.3213866}) noted that the defect might come from various sources, e.g., program code, execution environment, library framework itself, etc. In contrast, the focus of this study is to gain an understanding of ML API contracts.~\cite{islam19} reported on API misuse. API misuse can be 
detected if contract obligations are specified. \cite{8305957} investigated the issues in various ML libraries to understand the bug-fix patterns in these libraries, whereas \cite{islam20repairing} studied the deep neural network (DNN) models to understand the bug-fix patterns. In our study, we focused on ML API contracts and corresponding breaches. Suppose a user maintains a contract obligation for an ML API. In that case, if the API demonstrates exceptional behavior upon exiting, the issue may be present in the implementation of the API.

Our work focuses on investigating the kinds of contracts required to establish 
the correct usage of ML APIs. The main question is: 
{\em what are the kinds of contracts required to establish the correct usage of 
ML APIs?} 
We observe that
\ML~software is different from traditional software in several ways. 
In \ML~software, problem-solving is largely dependent on training 
data and subject to precise settings of hyper-parameters \cite{10.1145/3213846.3213866}. 
A prior work by \cite{dfaults} suggested that choice of loss function/optimizer, 
missing/redundant/wrong layers, etc. are distinctive bugs in \ML~software. 
Also, incorrect use of \ML~APIs may not always lead to crashes, but may instead lead to 
slower performance or statistically invalid results. In this study, we did not aim to check the reliability of the \ML~systems. Instead, we looked at the errors occurring in \ML~programs due to the incorrect usage of \ML~APIs.

We studied four popular \ML~libraries: 
\tf,~\scikit,~\keras and~\torch~and studied posts from the Q\&A forum 
\emph{Stack Overflow} (\SO)~that contain one of these libraries in a tag.
The dataset (labeled \SO posts, queries, source codes, etc.) generated during our study are available in the \emph {figshare} repository, \url{https://figshare.com/s/c288c02598a417a434df}. 
This dataset includes a total of 1565 posts, from which we manually curated posts that 
hold 413  contracts for relevant \ML~APIs. 
We use this data to answer the following research questions:
 
\noindent
{\textbf{RQ1 (Root Cause and Effect):}} What are the root causes and effects behind \ML~contract violations? \\
{\textbf{RQ2 (Patterns):}}  Are there common patterns of \ML~contract violations? 
\\
{\textbf{RQ3 (Contract Comprehension Challenges):}}  When does understanding \ML contracts require an advanced level of \ML software expertise? \\
{\textbf{RQ4 (Contract Violation Detection):}}  Can checking contracts at the API level help detect the violation in early \ML~pipeline stages?

These questions, and the data that support their answers, help to answer the main question, i.e., they enable researchers and practitioners to pinpoint where immediate support is required in terms of contracts for \ML~APIs. The key findings from our study are summarized in Table~\ref{tab:claim}.
\begin{table}[htbp!]
	\centering
	\caption{Findings and Insights}
	\setlength{\tabcolsep}{3.8pt}
	\scriptsize
	\label{tab:claim}
	\begin{tabular}{|l|p{52mm}||p{52mm}||}
		\hline
		{\cellcolor{Gray} \bfseries RQ} & {\cellcolor{Gray} \bfseries Findings} & {\cellcolor{Gray} \bfseries Actionable Insight} \\ \hline
		RQ1 & Most frequent contracts for \ML~APIs: (\S{\ref{par:IC-1} })
		\begin{enumerate}[leftmargin=*]
			\item Constraint check on single arguments of an API.
			\item Order of API calls that become a requirement eventually.
		\end{enumerate}  & This is a good news because 
		the software engineering (SE) community can employ some existing contract mining approaches 
		to also mine contracts for ML APIs; but there might be a need to combine
		behavioral and temporal contract mining approaches that have 
		been independently developed thus far. \\ \hline
			RQ4 & \ML~API contracts that are commonly violated occur in earlier ML pipeline stages (\S{\ref{par:eps}}). & 
		A verification system with \ML~contract knowledge can explain whether a bug 
		in the \ML~system that used those APIs stemmed from an API contract breach. \\ \hline %for those APIs. \\ \hline
		RQ3 & The absence of precise error messages (\S{\ref{par:Error}}) due to system failures 
		makes contract comprehension and violation detection more challenging. & 
		As domain experts can understand the challenging ML contracts (\S{\ref{sec:dcc}}), 
		this knowledge encoded as contracts can enable improved debugging mechanisms. \\ \hline
		RQ1 & \ML~APIs require several type checking contracts specific to \ML~(\S{\ref{par:ML}}) 
		and inter-dependency (Table \ref{tab:socontract1}) between behavioral and temporal contracts. & 
		Programming methodology and tools for design by contract
		should include sufficient expressiveness for these additional types of contracts seen in \ML~APIs. \\ \hline
	\end{tabular}
\end{table}

The contributions of our paper are the following. 
We provide a taxonomy for \ML~API contracts and corresponding root causes. 
This taxonomy (\S{\ref{subsec:con}}) added five new leaf node categories of 
contracts (with respect to the leaf categories observed in traditional behavioral and temporal contracts) observed in our study. 
The work also identified the stages of ML pipelines in which the violations occur (API contract violation locations) or affect the software and presented a dedicated 
classification (\S{\ref{subsec:vl}}). 
To our knowledge, this is the first work that attempts to understand the types 
of required contracts needed to prevent problems that may arise when using these \ML~APIs in 
software systems. 
In \S{\ref{sec:result}}, in addition to answering the research questions, 
we analyze the outcomes related to contract breaches. 
Finally, we provide recommendations to researchers, consumers, and producers of \ML~APIs based on the findings. 

\section{Methodology}
\subsection{Overview}
\label{sec:data}
\begin{figure}[h]
    \includegraphics[width=\linewidth,scale=0.6]{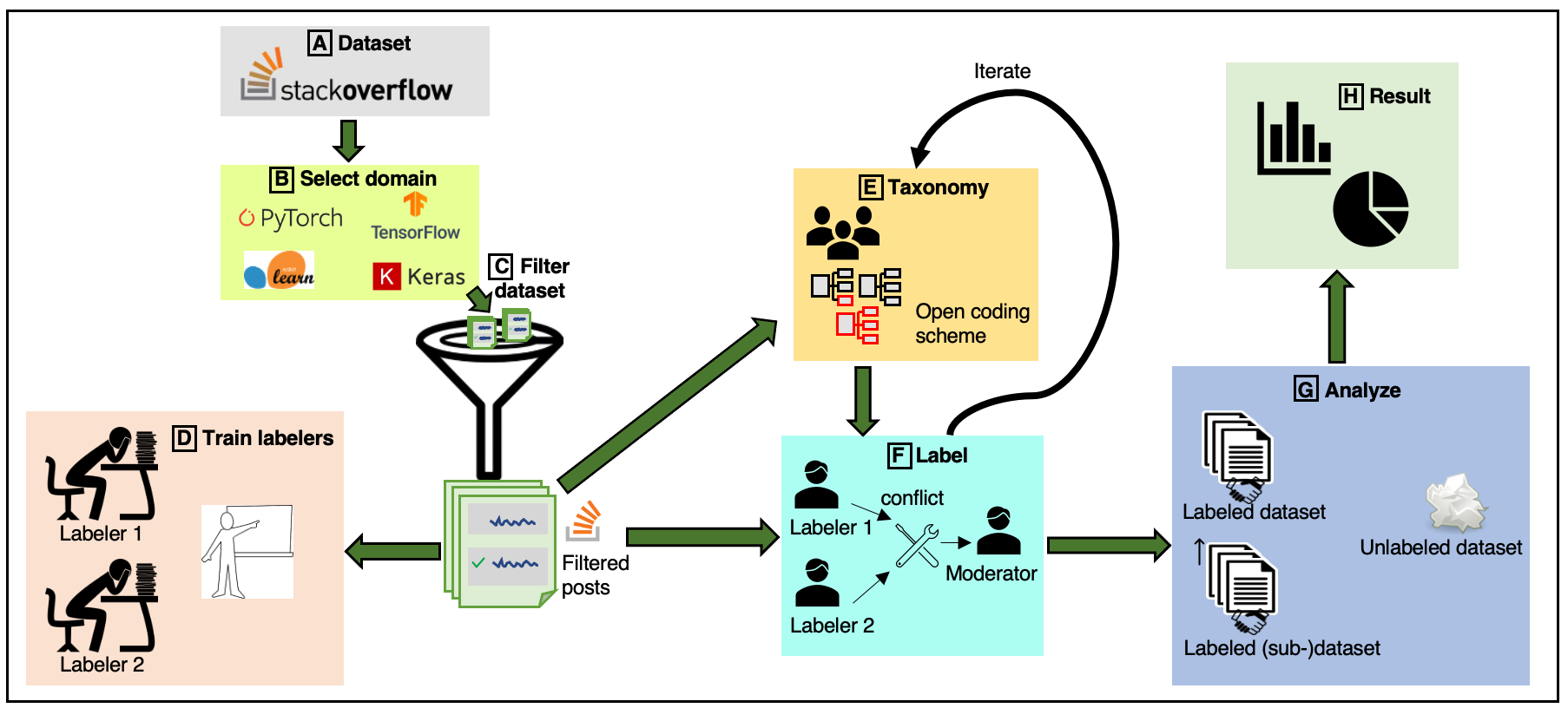}
    \caption{Overview of the adopted methodology}   
    \label{fig:overview}
\end{figure}

\boxed{A} In this study, we used
Stack Overflow (\SO) 
to investigate API contracts' requirements 
for the most-asked about and widely-used \ML~libraries and frameworks.
\SO is a forum for software development professionals and enthusiasts.
In recent years \SO~has served as an open repository for conducting studies on 
software engineering topics \cite{10.1145/3180155.3180260, 10.1145/3338906.3341186, Aghajani2019SoftwareDI, Beyer2014AMC, Barua2012WhatAD, Rosen2015WhatAM, Cummaudo2020InterpretingCC}. \SO, as a forum, maintains a strict moderation policy, promotes a peer-reviewing mechanism, and incorporates a reward system for encouraging quality answers from the software developers~\cite{reputationSO}.
Moreover, it has a vibrant user community and includes software developers from all walks of life, experiences, etc.~\cite{SOsurvey}.
As a result, it offers a wealth of well-vetted information on numerous software development topics. As such, \SO makes an excellent source for our study, as the primary goal of this study is to derive ML contracts from peer-reviewed and well-vetted content for the reliability of the findings.
To capture the contracts, analyzing the large code corpus of API usages~\cite{10.1145/1287624.1287632, 10.1145/1595696.1595767, khairunnesa2017exploiting} or the implementation of the software itself~\cite{cousot2013automatic} are both well-known techniques. Our chosen methodology is closer to the former.

\boxed{B} We used the SO forum's tags to identify the relevancy of a post to an ML library; if the question's tag contained an ML library name, it was considered a post related to that library and was thus a candidate to be studied in this work.

We ranked the top \ML~libraries using the \emph{frequency of these tags},
resulting in these four as the object of our study:
\tf, \scikit, \keras, and \torch.

\boxed{C} Next we filtered these posts based on a set of defined criteria that are described in detail in \S{\ref{ssec:filterdataset}}. 

\boxed{D} The second and third authors (labelers), both with a strong background in~\ML, were given background information on contract literature. Then they were given hands-on training with sample SO posts as described in \S{\ref{subsec:label}}. 

\boxed{E} After the training process, 10\% of the filtered dataset is used by the first three authors to develop the taxonomies used to label the filtered posts. Two iterations were needed to propose the final taxonomy presented here. The process is described in detail in \S{\ref{subsec:class}}. 

\boxed{F} Next, these labelers identified contracts implicitly present in \SO~posts. We obtained 162, 122, 103, and 26 contracts, respectively, from the previously curated posts. Table \ref{tab:datasummary} shows a summary of the dataset for each library in our study. For each \SO~question, we used the taxonomy of contracts (including proposed categories)  
from \S{\ref{subsec:con}} to investigate the available information from the question 
and accepted answer to decide the type of contract obligation missing in the question 
and marked in the response. Hence, if the~\SO response describes the correct way of using an API of interest violated in the question, we identify that as a contract for the API that was implicitly present in~\SO posts. We have also used the taxonomies presented in \S{\ref{subsec:vl}} and \S{\ref{subsec:class}} to complete the labeling in this stage. In \S{\ref{subsec:con}}, we describe the process of identifying the contract violation and potential contract for \ML APIs with example \SO posts from our study. Then, the first author, with expertise (of approx. $6$ years) in contracts, reviewed the identified contracts and the~\SO post the contracts were extracted from. This served two purposes: it ensured that the identified contracts were correct and helped to reduce the threat of missing contracts that was implicitly present in the dataset from the second and third authors. The first author found only a handful of contracts missed by these two labelers. However, these missing contracts were found by at least one of the two labelers. Therefore, we did not note any new contracts this way. If one labeler identified a contract and the other did not, as they performed their labeling using the proposed taxonomy, this was identified as one of the reasons behind creating a conflict between the two labelers. We discussed in \S{\ref{subsec:label}} how we resolved the conflicts in our study.

\boxed{G} As the labeling process is completed, we analyze our labeled dataset. Additionally, we have created a separately filtered dataset (a subset of the original) based on the question scores and analyzed questions with a relatively high score (in the range of 30-339). The intuition behind further separating these posts is that an author may ask one question, and only a handful of \ML API users might run into it. Then another \SO~question may be inquired by someone but up-voted by hundreds of others who have the same problem. Thus, the intuition behind further separating these filtered posts was to understand how many \ML API users are struggling with each problem. This separate dataset was compared against the entire dataset to be vigilant about the representative issues and respective conclusions we draw from the posts in \S{\ref{sec:result}.} This subset is selected with the following criteria: select high-quality posts and keep manual efforts manageable. To that end, to ensure high-quality posts, we select posts having better than average scores (avg. 18.9). To keep the manual effort manageable, we find a trade-off between sample size and its statistical power. We specifically choose 30 as a cut-off to have reasonable confidence in this additional study while keeping manual efforts manageable (about 90\% confidence level with a 5\% margin of error), resulting in 222 posts. 
We discarded posts if they did not capture any information regarding correct usage of \ML~APIs. 
Additionally, we grouped the discarded \SO~posts that we could  not label as containing contracts 
during the manual curation step. 
We were unable to label some posts due to the following reasons: posts asking general 
clarification questions, unresolved issues with specific APIs of interest, an API unidentified 
in a post, a solution involving tuning, or a dependency between an unrelated API and 
a related API. For instance, in some of these posts, the \ML API users is usually curious about performing a task at hand or inter-library code transformation and refactoring. To illustrate, in one \SO~post\footnote{https://stackoverflow.com/questions/42480111/}, the author is knowledgeable that they can use the \texttt{summary()} API from Keras to print a model summary. Yet, they want to know how to do the same using Pytorch. Even if these posts are of interest to \ML API users, these do not fall into the category of contract requirements. In our study, the number of total unlabelled posts is 1159, and the number of total unlabelled posts with a score of or higher 30  is 161.

\boxed{H} Finally we present our result in detail in \S{\ref{sec:result}}. However, we did not add the statistics for unlabeled posts in RQs, as these posts did not unveil any \ML contracts. 

\begin{table}[htbp!]
\centering
	\caption{Dataset for Empirical Study on ML Contracts}
\begin{tabular}{|c|rrrr|r|}
\hline
\rowcolor[HTML]{C0C0C0} 
\cellcolor[HTML]{C0C0C0} & \multicolumn{4}{c|}{\cellcolor[HTML]{C0C0C0}SO \# Posts}                                  & \multicolumn{1}{c|}{\cellcolor[HTML]{C0C0C0}} \\ \cline{2-5}
\rowcolor[HTML]{EFEFEF} 
\multirow{-2}{*}{\cellcolor[HTML]{C0C0C0}Library} &
  \multicolumn{1}{c|}{\cellcolor[HTML]{EFEFEF}Criteria \boxed{1.}} &
  \multicolumn{1}{c|}{\cellcolor[HTML]{EFEFEF}Criteria \boxed{2.}} &
  \multicolumn{1}{c|}{\cellcolor[HTML]{EFEFEF}Criteria \boxed{3.}} &
  \multicolumn{1}{c|}{\cellcolor[HTML]{EFEFEF}Criteria \boxed{4.}} &
  \multicolumn{1}{c|}{\multirow{-2}{*}{\cellcolor[HTML]{C0C0C0}\# Contracts}} \\ \hline
\tf       & \multicolumn{1}{r|}{24368} & \multicolumn{1}{r|}{2205} & \multicolumn{1}{r|}{1400} & 605  & 162                                           \\ \hline
\scikit   & \multicolumn{1}{r|}{12506} & \multicolumn{1}{r|}{1641} & \multicolumn{1}{r|}{1127} & 551  & 122                                           \\ \hline
\keras    & \multicolumn{1}{r|}{12300} & \multicolumn{1}{r|}{1285} & \multicolumn{1}{r|}{821}  & 333  & 103                                           \\ \hline
\torch    & \multicolumn{1}{r|}{2500}  & \multicolumn{1}{r|}{439}  & \multicolumn{1}{r|}{313}  & 76   & 26                                            \\ \hline
Total                    & \multicolumn{1}{r|}{51674} & \multicolumn{1}{r|}{5570} & \multicolumn{1}{r|}{3661} & {\cellcolor[HTML]{EFEFEF}1565} & {\cellcolor[HTML]{EFEFEF}413}                                           \\ \hline
\end{tabular}%
\label{tab:datasummary}
%}
\end{table}
\subsection{Filter Dataset} 
\label{ssec:filterdataset}

We processed the collected posts further to enable a classification scheme for contracts. We followed two main steps to filter these posts. The \emph{initial step} is an \emph{automatic} pre-processing of the collected posts 
based on the following criteria: 
\boxed{1.} Within these posts, authors asking the question must include some code snippet(s). 
We reason that a question post discussing these libraries and including snippets 
of code is more likely to have difficulty with API contracts, thus
may show challenges related to contract violation for relevant APIs. %\textcolor{blue}{
Furthermore, posts with code enable identifying the \ML~APIs being used. We have collected a total of 51674 posts with this filtering criteria. \boxed{2.} We further filter the posts having a score higher than five based on 
the guidelines from prior works~\cite{6405249, 10.1145/3213846.3213866, islam19, islam20repairing} to 
ensure that posts are of high quality. This additional criterion produced a total of 5570 posts.
\boxed{3.} We considered posts with accepted answers (having a score higher than five) only, as those answer 
posts have successfully identified and resolved the problem faced 
by the author of the question post. The criterion for including accepted answers to the dataset follows prior studies~\cite{islam19, 8987482, islam20repairing} that have argued that if the answer of a post is not accepted, then that answer might not have addressed the issue. This step enabled us to collect 3661 posts in total. The steps up to this point are automatic. All queries to filter datasets according to this set of criteria are provided in the \emph{figshare} repository\footnote{\url{https://figshare.com/s/c288c02598a417a434df}}.

\boxed{4.} Furthermore, the accepted answers frequently contain code, and 
we expect that these code snippets focus on the required contracts for \ML~APIs. 

Additionally, we manually checked if the accepted answer to a post clearly describes the API contracts using text (without code). If this is true, we also consider such a post. The question posts in our study provided us with the contract violations \ML~software is susceptible to and the accepted answer posts directed us towards the contracts. Thus, the considered \SO~posts capture both contract violations and potential contracts. After these stages, we curated a total of 1565 posts; the posts
specific to each of \tf,~\scikit,~\keras,~\torch~contained 605, 551, 333, and 76 posts (Table \ref{tab:datasummary}), respectively. 

The posts obtained after manual filtering each contain at least one \ML API-related contract but may contain more. Our study observed a blend of behavioral and temporal contracts for \ML APIs. We called this a hybrid category in our classification (in \S{\ref{subsec:con}}). The posts from where we have extracted such contracts are the source behind multiple contracts from a single post. If multiple contracts were present in a single post, we extracted all contracts and labeled these using our taxonomy. For instance, in one SO post (\SO post~\ref{fig:post1}), it contained two contracts. The API in question is \texttt{random\_shuffle()} from the TensorFlow library. The first extracted contract is to specify the argument seed with the desired value. The second extracted contract is to call \texttt{tf.random\_shuffle()} and then call \texttt{tf.reset\_default\_graph()}. And the \texttt{random\_shuffle()} API will ensure a shuffled \texttt{Tensor} if the contract is maintained in either of the ways mentioned.

Besides, the \SO~forum has a general strategy to tackle duplicate questions with the same (potential) answers. Users can flag a question on~\SO if it is analogous to a previously posted question concerning a concept. According to~\SO, the reasoning behind marking duplicate questions is that users should not discuss duplicate questions, but anyone with the same query can refer to the previously posted discussions. In our study, we have found 359 unique contracts from a total of 413 contracts reported. SO’s strategy for flagging duplicating questions has enabled us to collect many unique ML contracts.

We note that the forum may contain other relevant \ML API posts but not included in our dataset 
if the posts do not contain any contract or match the filtering criteria mentioned above. 
We have inspected the impact of imbalance in our dataset across libraries and address this in \S{\ref{ssec:threat}}. 

Next, we present a classification for \ML~API contracts and associated root causes 
in \S{\ref{subsec:con}}. 
This classification is used to identify and label posts with \ML~contracts. \S{\ref{subsec:class}} demonstrates the taxonomy of the effects of these root causes of contract violations. Finally, we present a classification to identify locations of \ML~API contract violation (\S{\ref{subsec:vl}}) based on \ML~pipeline stages.

\subsection{Classification of \ML~Contracts and Violation Root Causes}
\label{subsec:con}

To label the contracts for \ML~APIs found in our dataset, we developed a classification 
scheme that categorizes different types of contracts originating from these APIs.

As mentioned earlier, the literature mainly discusses two types of contracts: behavioral and temporal. 

Typically, behavioral contracts for APIs consist of assertions that are required to be true before calling the API (preconditions) and assertions that must be valid upon exiting the API method (postconditions).
In contrast, temporal contracts are those that capture the required 
order of API calls to ensure proper behavior. Both types of contracts are also observed in non-ML APIs, and we build our classification on top of this well-established classification. Building on an existing classification scheme helped us to not reinvent known ideas \cite{10028142446} related to API contracts. Student authors in this work used open coding to build the extension appropriate for \ML~APIs. 

{\bf Process:\ } Researchers advocate using open coding to
create any taxonomy~\cite{926934}; it is best that the researchers perform the
task themselves rather than rely on a third party. The authors worked as a group 
initially to perform the coding and sampled 10\% data to that purpose. 
This strategy had several advantages, e.g., a consistent decision to choose between existing concepts and create a new one; categories became more exact while differences became more evident than individually proposed taxonomy categories, and it also provided an opportunity to properly train the two labelers. We used \emph{axial coding} \cite{Corbin2008BasicsOQ}, a technique that helps to collapse core themes involving qualitative data. In other words, it organizes the codes developed during open coding. This technique is used for cases where conceiving sub-categories seems necessary for any central component inside the classification schema. To elaborate, in our study, as we analyzed and labeled the SO posts with identified contracts, we looked at how these sub-categories could be grouped into central categories, so that the central category could encompass a number of different posts. In some cases, these central categories (axes) are from the state-of-the-art taxonomy, e.g., data type-related contracts, but in other cases, a new abstract category seemed appropriate, e.g., selection. For instance, the codes such as \emph{Primitive Type}, \emph{Built-in Type}, etc., are well-established codes that describe different categories of type-related contracts. We used axial coding to identify that these contracts can be collapsed into the sub-core theme of checking \emph{Data Type}-related contracts. Similarly, we organized sub-core core categories eventually into core categories. For instance, \emph{Data Type} is organized under the core category \emph{Single API Method}.
We further use relational and variational sampling~\cite{GroundedTheoryResearchProceduresCanonsandEvaluativeCriteria} 
using \SO~data to support or contradict the relationship between 
sub-categories and core categories. 
These sampling techniques facilitated explaining relations between theoretically 
relevant categories through gathering data (depending on the frequency of 
similarity or variation) on each group, e.g., considering conditions, 
consequences, etc., on a case-by-case basis. For example, we located instances of the leaf category \emph{\ML type} in our dataset that describes special type-related contracts that is only present in \ML APIs. The multiple samples we collected indicated that the reason behind this contract violation is the \emph{input of an unacceptable input type}, and the effect, if explicitly present in the samples, is \emph{crash}. The frequency of such similarity confirmed the relationship between the category \emph{\ML type} and the category \emph{Data Type}. This is an example of relational sampling, precisely.

{\bf New Categories:\ } We found four new categories during our initial study (marked with $\tikzcircle[fill=olive]{3pt}$ in Table \ref{tab:class1}). 
After the initial study, the labelers individually 
studied the rest of the posts and were at liberty to suggest additional categories if 
the need arose (detail on labeling in \S{\ref{subsec:label}}). 
The labelers conducted an in-person meeting under the supervision of a moderator 
to discuss the suggested additional categories and these reconciliation effort resulted 
in one additional category (marked with %$\Diamond$ 
${\LARGE \tikztriangleright[red,fill=orange!50]}$ in Table \ref{tab:class1}).

{\bf Classification Scheme:\ }
Next, we described our obtained classification schema in detail. 
Furthermore, all categories included in this classification are shown in 
Table \ref*{tab:class1}. 
At the top level, 
we presented three central contract component levels: contracts involving \emph{Single API Method}, 
contracts involving \emph{API Method Order}, and contracts that 
required a \emph{Hybrid} of preceding categories. The first fundamental category, \emph{Single API Method (SAM)}, in our classification scheme captures preconditions/postconditions involving a single API method. This core category is based on behavioral contracts. Next, \ML~APIs often require particular call orderings to demonstrate normal behavior;  we classify contracts specifying such order as {\bfseries API Method Order (AMO)}. This category is based on temporal contracts.
Subsequently, we classified these categories into sub-classes until we could find a leaf category that denoted the contract of a particular type for \ML~APIs. For each such class, we explained the root cause of that contract violation subsequently.

\begin{table}[]
    \centering
	\captionof{table}{Type of Contracts for \ML APIs (Symbols ${\tikzcircle[fill=olive]{3pt}}$ and ${\LARGE \tikztriangleright[red,fill=orange!50]}$ at the end of leaf components designate novel categories)}
	\label{tab:class1}
\resizebox{\textwidth}{!}{%
\begin{tabular}{|c|cl|}
\hline
\rowcolor[HTML]{EFEFEF} 
\multicolumn{1}{|l|}{\cellcolor[HTML]{EFEFEF}Level 1} & \multicolumn{1}{l|}{\cellcolor[HTML]{EFEFEF}Level 2} & Level 3 \\ \hline
\rowcolor[HTML]{32CB00} 
\cellcolor[HTML]{32CB00} & \multicolumn{1}{c|}{\cellcolor[HTML]{32CB00}} & Primitive Type (PT) \\ \cline{3-3} 
\rowcolor[HTML]{32CB00} 
\cellcolor[HTML]{32CB00} & \multicolumn{1}{c|}{\cellcolor[HTML]{32CB00}} & Built-in Type (BIT) \\ \cline{3-3} 
\rowcolor[HTML]{32CB00} 
\cellcolor[HTML]{32CB00} & \multicolumn{1}{c|}{\cellcolor[HTML]{32CB00}} & Reference Type (RT) \\ \cline{3-3} 
\rowcolor[HTML]{32CB00} 
\cellcolor[HTML]{32CB00} & \multicolumn{1}{l|}{\multirow{-4}{*}{\cellcolor[HTML]{32CB00}Data Type (DT)}} & ML Type (MT) ${\tikzcircle[fill=olive]{3pt}}$ \\ \cline{2-3} 
\rowcolor[HTML]{32CB00} 
\multirow{-5}{*}{\cellcolor[HTML]{32CB00}Single API Method (SAM)} & \multicolumn{1}{c|}{\cellcolor[HTML]{32CB00}} & Intra-argument Contract (IC-1) \\ \cline{3-3} 
\rowcolor[HTML]{32CB00} 
\multicolumn{1}{|l|}{\cellcolor[HTML]{32CB00}} & \multicolumn{1}{l|}{\multirow{-2}{*}{\cellcolor[HTML]{32CB00}Boolean Expression Type (BET)}} & Inter-argument Contract (IC-2) \\ \hline
\rowcolor[HTML]{68CBD0} 
\cellcolor[HTML]{68CBD0} & \multicolumn{2}{l|}{\cellcolor[HTML]{68CBD0}Always (G)} \\ \cline{2-3} 
\rowcolor[HTML]{68CBD0} 
\multirow{-2}{*}{\cellcolor[HTML]{68CBD0}API Method Order (AMO)} & \multicolumn{2}{l|}{\cellcolor[HTML]{68CBD0}Eventually (F)} \\ \hline
\rowcolor[HTML]{FFC702} 
\multicolumn{1}{|l|}{\cellcolor[HTML]{FFC702}} & \multicolumn{1}{l|}{\cellcolor[HTML]{FFC702}SAM-AMO Interdependency (SAI)} & SAM (Level 3) $\land$ AMO(Level 2) ${\tikzcircle[fill=olive]{3pt}}$ \\ \cline{2-3} 
\rowcolor[HTML]{FFC702} 
\multicolumn{1}{|l|}{\cellcolor[HTML]{FFC702}} & \multicolumn{1}{c|}{\cellcolor[HTML]{FFC702}} & SAM (Level 3) ${\LARGE \tikztriangleright[red,fill=orange!50]}$ \\ \cline{3-3} 
\rowcolor[HTML]{FFC702} 
\multicolumn{1}{|l|}{\cellcolor[HTML]{FFC702}} & \multicolumn{1}{c|}{\cellcolor[HTML]{FFC702}} & AMO (Level 2) ${\tikzcircle[fill=olive]{3pt}}$ \\ \cline{3-3} 
\rowcolor[HTML]{FFC702} 
\multicolumn{1}{|l|}{\multirow{-4}{*}{\cellcolor[HTML]{FFC702}Hybrid (H)}} & \multicolumn{1}{l|}{\multirow{-3}{*}{\cellcolor[HTML]{FFC702}Selection (SL)}} & Comb. of SAM(Level 3) and AMO(Level 2) $\tikzcircle[fill=olive]{3pt}$ \\ \cline{3-3} 
\end{tabular}%
}
	\footnotesize * %\textbf{\textcolor{green}{Green}}
		\colorbox{lightgreen}{Green} cells indicates the behavioral contract. \colorbox{lightblue}{Blue} denotes temporal contract and \colorbox{lightorange}{Orange} cells indicate the hybrid respectively.
\end{table}

\subsubsection{Type of Contracts involving Single API Method (SAM)}

The first sub-category of Single API Method (SAM) contract concerns type checking 
that is required {\bfseries Data Type (DT)} of API arguments.

This subclass consists of four types of contract: 
\begin{description}[leftmargin=*]
	\item[\bfseries Primitive Type (PT):] This represents the \ML~API argument type 

can be a primitive type, e.g., \texttt{float},  \texttt{int},  \texttt{bool},  \texttt{number},  \texttt{None}, and the rest. For instance, in \SO post \ref{fig:postPT}, the \texttt{decode()} method from the \tf library expects a \texttt{byte string}. The root cause of this contract violation is an input of an unacceptable type.  

\hspace*{0.2cm}
{
	\begin{lrbox}{\mybox}%
		\begin{lstlisting}[language=iPython]
hello = tf.constant('Hello, TensorFlow!')
sess = tf.Session()
print(sess.run(hello))
\end{lstlisting}
	\end{lrbox}%
	\scalebox{0.92}{\usebox{\mybox}}
	\begin{mytbox2}[colback=green!15,fontupper=\scriptsize,flushleft upper,boxrule=0pt,arc=5pt,left=2pt,right=2pt,after=\ignorespacesafterend\par\noindent]{\checkmark Accepted Answer}
		Use sess.run(hello).decode() because it is a bytestring, decode method will return the string.
	\end{mytbox2}
	\captionof{Stack Overflow post}{\href{https://stackoverflow.com/questions/40904979/the-print-of-string-constant-is-always-attached-with-b-intensorflow}{Example post with contract} \label{fig:postPT}}}
\hspace*{0.2cm}
{
\begin{lrbox}{\mybox}%
 \begin{lstlisting}[language=iPython]
conv2 = conv2d(conv1, wts['conv2'])
conv2 = maxpool2d(conv2)
conv2 = tf.reshape(conv2, shape=[-1,158*117*64])
frame = tf.placeholder('float', [None, 640-10, 465, 3])
controls_at_each_frame = tf.placeholder('float', [None, 4]) 
conv2 = tf.concat(conv2, controls_at_each_frame, axis=1)
\end{lstlisting}
\end{lrbox}%
\scalebox{0.92}{\usebox{\mybox}}
	
	\begin{mytbox2}[colback=green!15,fontupper=\scriptsize,flushleft upper,boxrule=0pt,arc=5pt,left=2pt,right=2pt,after=\ignorespacesafterend\par\noindent]{\checkmark Accepted Answer}
%	\begin{lstlisting}[language=iPython]
    \texttt{conv2 = tf.concat((conv2, controls\_at\_each\_frame), axis=1)} ;
   % \end{lstlisting}
Note we need two frames that are required to concatenate within parentheses, as specified
	\end{mytbox2}
	\captionof{Stack Overflow post}{\href{https://stackoverflow.com/questions/45175469/typeerror-concat-got-multiple-values-for-argument-axis}{Example post with contract} \label{fig:postBIT}}}

\item[\bfseries Built-in Type (BIT):] The contracts involving more complex 
built-in types (such as \texttt{dict}, \texttt{list}, \texttt{tuple}, and \texttt{array}). For example, in \SO post \ref{fig:postBIT}, \texttt{concat()} from the \tf library expects the first argument to be of \texttt{array} type. The root cause of this contract violation is an input of unacceptable type.
\end{description}

\begin{description}[leftmargin=*]
	\item[\bfseries Reference Type (RT):] This category of   contracts can involve either internal class object, i.e., referenced class objects within the API class, or external class object, i.e., external variable referenced from separate modules of the ML library. For example, in \SO post \ref{fig:post3}, a contract for the API \texttt{KerasRegressor()} from \keras is shown. The argument accepts a function, an instance of a class that implements the call method or \texttt{None}. As the argument \texttt{build\_fn} of this API accepts reference type as one of its expected argument types, we classify this under the reference type category. 
 The root cause of this contract violation is that an input of unacceptable  type is supplied to the method. 
\end{description}

{
	\begin{lrbox}{\mybox}%
		\begin{lstlisting}[language=iPython]
model = nn_model()
model = KerasRegressor(build_fn=model, nb_epoch=2) \end{lstlisting}
	\end{lrbox}%
	\scalebox{0.92}{\usebox{\mybox}}
	
	\begin{mytbox2}[colback=green!15,fontupper=\scriptsize,flushleft upper,boxrule=0pt,arc=5pt,left=2pt,right=2pt,after=\ignorespacesafterend\par\noindent]{\checkmark Accepted Answer}
		Herein lies the issue. Rather than passing your \texttt{nn\_model} function as the \texttt{build\_fn}, you pass an actual instance of the \texttt{Keras} Sequential model. One of the following three values could be passed to \texttt{build\_fn}: \newline
		\textbullet~A function \newline
		\textbullet~An instance of a class that implements the call method. \newline
		\textbullet~None
	\end{mytbox2}
	\captionof{Stack Overflow post}{\href{https://stackoverflow.com/questions/39467496/error-when-using-keras-sk-learn-api}{Example post with contract} \label{fig:post3}}}
\hspace*{0.2cm}

\begin{description}[leftmargin=*]
	\item[\bfseries ML Type (MT):] This final contract component of data type 
contains \ML types. ML types are a multidimensional array with a uniform type (\texttt{float16}, \texttt{float32}, \texttt{complex16}, etc.), particularly designed for ML pipelines to achieve accelerated performance (i.e., ease of use with GPU).

For instance, in \SO post~\ref{fig:post2}, an \emph{ML Type} 
related contract is captured stating that the \texttt{matmul()} API from the \tf library 
requires that both of the arguments should be a \texttt{Tensor} with one of the following types: 
\texttt{float16}, \texttt{float32}, \texttt{float64}, \texttt{int32}, \texttt{complex64}, \texttt{complex128}. \\

\hspace*{0.2cm}
{
	\begin{lrbox}{\mybox}%
		\begin{lstlisting}[language=iPython]
layer_1 = tf.nn.relu(tf.add(tf.matmul(_X, _weights['h1']), _biases['b1'])) 
layer_2 = tf.nn.relu(tf.add(tf.matmul(layer_1, _weights['h2']), _biases['b2'])) 
return tf.matmul(layer_2, weights['out']) + biases['out'] 
\end{lstlisting}
	\end{lrbox}%
	\scalebox{0.92}{\usebox{\mybox}}
\begin{mytbox2}[colback=green!15,fontupper=\scriptsize,flushleft upper,boxrule=0pt,arc=5pt,left=2pt,right=2pt,after=\ignorespacesafterend\par\noindent]{\checkmark Accepted Answer}
	The \texttt{tf.matmul()} op does not perform automatic type conversions, so both of its inputs must have the same element type. The error message you are seeing indicates that you have a call to \texttt{tf.matmul()} where the first argument has type \texttt{tf.float32}, and the second argument has type \texttt{tf.float64}. You must convert one of the inputs to match the other, for example using \texttt{tf.cast(x, tf.float32)}.
\end{mytbox2} 
\captionof{Stack Overflow post}{\href{https://stackoverflow.com/questions/36210887/how-to-fix-matmul-op-has-type-float64-that-does-not-match-type-float32-typeerror}{Example post with contract} \label{fig:post2}}
}
Another example of this type of contract \ref{fig:postMT} 
is that the first two arguments for the  \texttt{fit()} API should have the type of a \texttt{numpy} \texttt{array} or a \texttt{list} of  \texttt{numpy}  \texttt{arrays}. The root cause of this contract violation is an input of unacceptable type supplied to the method. 
This post also shows that the API has a supplementary contract concerning argument dependency, as discussed below. \\

\hspace*{0.2cm}
{
	\begin{lrbox}{\mybox}%
		\begin{lstlisting}[language=iPython]
padded_model.fit(train_X, train_y, epochs=50, verbose=1)
\end{lstlisting}
	\end{lrbox}%
	\scalebox{0.92}{\usebox{\mybox}}
	
	\begin{mytbox2}[colback=green!15,fontupper=\scriptsize,flushleft upper,boxrule=0pt,arc=5pt,left=2pt,right=2pt,after=\ignorespacesafterend\par\noindent]{\checkmark Accepted Answer}
If \texttt{train\_x} and \texttt{train\_y} are normal \textit{Python} lists, they don't have the attribute .ndim. Only Numpy arrays have this attribute representing the number of dimensions.
	\end{mytbox2}
	\captionof{Stack Overflow post}{\href{https://stackoverflow.com/questions/48699954/keras-attributeerror-int-object-has-no-attribute-ndim-when-using-model-fi}{Example post with contract} \label{fig:postMT}}}

\end{description}

The API method can also involve Boolean assertions related to its argument values,
{\bfseries Boolean Expression Type (BET)}, instead of only type related checks.
We classify these types of contracts into two subclasses: 

\begin{description}[leftmargin=*]
\item 
[\bfseries Intra-argument contracts (IC-1):] IC-1 specifies preconditions related to a single argument of the API. These contracts may involve both comparisons and logical combinations. 

{\quad\begin{lrbox}{\mybox}%
\begin{lstlisting}[language=iPython, numbers=none]
a_shuf = tf.random_shuffle(a)
\end{lstlisting}
	\end{lrbox}%
	\scalebox{0.92}{\usebox{\mybox}}
	
	\begin{mytbox2}[colback=green!15,fontupper=\scriptsize,flushleft upper,boxrule=0pt,arc=5pt,left=2pt,right=2pt,after=\ignorespacesafterend\par\noindent]{\checkmark Accepted Answer}
That only sets the graph-level random seed. If you execute this snippet several times in a row, the graph will change, and two shuffle statements will get different operation-level seeds.
		To get deterministic \texttt{a\_shuf} you can either \newline
		\textbullet~Call \texttt{tf.reset\_default\_graph()} between invocations or \newline
		\textbullet~Set operation-level seed for shuffle: \texttt{a\_shuf = tf.random\_shuffle(a, seed=42)}
	\end{mytbox2}
	\captionof{Stack Overflow post}{\href{https://stackoverflow.com/questions/36096386/tensorflow-set-random-seed-not-working}{Example post with contract} \label{fig:post1}}}
\hspace*{0.2cm}

An example of an IC-1 contract is given in \SO post~\ref{fig:post1}, which shows an \ML API users trying
to use the \tf~API \texttt{random\_shuffle()} to shuffle a Tensor, \texttt{a}, 
with some set seed value. 
One of the solutions mentioned in the accepted answer says that to do that, one should specify the argument \texttt{seed} with the desired value, e.g., the argument \texttt{seed} gets the value \texttt{42}. 
The root cause of this contract violation is that acceptable input value is not supplied to (the \texttt{random\_shuffle()}) method. 
\end{description}

\begin{description}[leftmargin=*]
\item [\bfseries Inter-argument contracts (IC-2):] 
IC-2 contracts involve more than one argument to an API method, possibly using comparisons or logical expressions.
For example, in \SO post~\ref{fig:post2}, the \texttt{matmul()} API from  \tf requires that
the type of the second argument should match the type of the first argument. A comparison expression can express this contract, so it belongs to IC-2. 
The root cause of this contract violation is that the (\texttt{matmul()}) API is missing input value/type dependency between arguments. Another example \ref{fig:postIC2}

for this category is \texttt{nn.softmax\_cross\_entropy\_with\_logits()}, an API from \tf, which requires that the logits and labels arguments must have the same shape (i.e., [batch\_size, num\_classes]). \\

\hspace*{0.2cm}
{
	\begin{lrbox}{\mybox}%
		\begin{lstlisting}[language=iPython]
tf.nn.softmax_cross_entropy_with_logits(
    logits=b, labels=a).eval(feed_dict={b:np.array([[0.45]]), a:np.array([[0.2]])})
\end{lstlisting}
	\end{lrbox}%
	\scalebox{0.92}{\usebox{\mybox}}
	
	\begin{mytbox2}[colback=green!15,fontupper=\scriptsize,flushleft upper,boxrule=0pt,arc=5pt,left=2pt,right=2pt,after=\ignorespacesafterend\par\noindent]{\checkmark Accepted Answer}
Like they say, you can't spell \texttt{"softmax\_cross\_entropy\_with\_logits"} without \texttt{"softmax"}. \texttt{Softmax} of [0.45] is [1], and log(1) is 0. \texttt{logits} and \texttt{labels} must have the same shape \texttt{[batch\_size, num\_classes]} and the same dtype (either \texttt{float16, float32, or float64}).
	\end{mytbox2}
	\captionof{Stack Overflow post}{\href{https://stackoverflow.com/questions/42521400/calculating-cross-entropy-in-tensorflow}{Example post with contract} \label{fig:postIC2}}}

\end{description}

\subsubsection{Type of Contracts involving API Method Order (AMO)}

Multiple APIs can be involved in an AMO contract. There are two sub-categories as follows:

\begin{description}[leftmargin=*]
	\item [\bfseries Always (G):] Always contracts are AMO contracts that hold at each point of history. %must hold in all program states. 
	For example, as shown in \SO post~\ref{fig:post5}, for \tf, the call to the method, \texttt{tf.wholeFileReader()} must be followed by another method call, \texttt{tf.train.start\_queue\_runners()} to avoid hanging. 
The root cause of this contract violation is that the \emph{always} required order between these calls is not followed. \\
\end{description}

{\begin{lrbox}{\mybox}%
		\begin{lstlisting}[language=iPython]
def read_image(filename_queue):
reader = tf.WholeFileReader()
key,value = reader.read(filename_queue) \end{lstlisting}
	\end{lrbox}%
	\scalebox{0.92}{\usebox{\mybox}}
	
	\begin{mytbox2}[colback=green!15,fontupper=\scriptsize,flushleft upper,boxrule=0pt,arc=5pt,left=2pt,right=2pt,after=\ignorespacesafterend\par\noindent]{\checkmark Accepted Answer}
		If you're not using feeding—i.e. using the \texttt{tf.WholeFileReader} as shown in your program—you will need to call \texttt{tf.train.start\_queue\_runners()} to get started. Otherwise your program will hang, waiting for input.
	\end{mytbox2}
	\captionof{Stack Overflow post}{\href{https://stackoverflow.com/questions/35673874/tensorflow-error-shape-tensorshape-must-have-rank-1}{Example post with contract} \label{fig:post5}}
	\hspace*{0.2cm}}

{\begin{lrbox}{\mybox}%
		\begin{lstlisting}[language=iPython]
model = Sequential()
model.add(LSTM(100, input_dim = num_features))
model.add(Dense(1, activation='sigmoid')) \end{lstlisting}
	\end{lrbox}%
	\scalebox{0.92}{\usebox{\mybox}}
	
	\begin{mytbox2}[colback=green!15,fontupper=\scriptsize,flushleft upper,boxrule=0pt,arc=5pt,left=2pt,right=2pt,after=\ignorespacesafterend\par\noindent]{\checkmark Accepted Answer}
The solution to this: you need to enable the \texttt{LSTM} layer to return a sequence instead of only the last element. Since the \texttt{Dense} layer is not able to handle sequential data you need to apply it to each sequence element individually which is done by wrapping it in a \texttt{TimeDistributed} wrapper.
	\end{mytbox2} 
	\captionof{Stack Overflow post}{\href{https://stackoverflow.com/questions/41863921/how-can-i-use-categorical-one-hot-labels-for-training-with-keras}{Example post with contract} \label{fig:post6}}}
	
\begin{description}[leftmargin=*]
	\item [\bfseries Eventually (F):] Eventually contracts are AMO contracts where the ordering is 
	only required at some point in history. In other words, this specifies that a required API ordering must be true at some point in this program's execution history far enough in the future. 
For instance, in \SO~post~\ref{fig:post6}, the author is trying to solve a sequential classification (input data where order matters) task. In the model, they used the \texttt{LSTM()} API to return a sequence and then output it as a \texttt{Dense()} object. The activation function has a one-to-one correspondence with the type of classification being performed. For that reason, the \texttt{sigmoid} function is rightly used. However, in the model, they used the \texttt{LSTM()} API to return a sequence and then output it as a \texttt{Dense()} object. This method order of APIs demonstrates an incorrect API method order, as the order in the question post is missing a \texttt{TimeDistributed()} API call.

Note that the code is correct for a many-to-one task in natural language processing (NLP). However, in this question, the user asks for a many-to-many solution, in which case it becomes mandatory to apply \texttt{TimeDistributed()}.
Therefore, using the \texttt{TimeDistributed()} API only becomes a requirement after the \texttt{LSTM()} API is used to return a sequence. The root cause of this contract violation is that in a state where a call to a method (\texttt{LSTM()}) returns (a sequence), another call to a method (\texttt{TimeDistributed()}) should have occurred. Thus, in this \SO~post~\ref{fig:post6}, this \emph{eventually} contract is violated because the author did not know that the \texttt{TimeDistributed()} API is a requirement to be called eventually after the \texttt{LSTM()} API is used to \emph{return a sequence}. \\

\end{description}

\subsubsection{Type of Contracts involving Hybrid (H) of SAM and AMO}
The {\bfseries  Hybrid (H)} category involves a blend of behavioral and temporal contracts. This category has two subclasses: \newline 
	 
\begin{lrbox}{\mybox}%
\begin{lstlisting}[language=iPython,numbers=none]
clf = GridSearchCV(SVC(C=1), tuned_parameters, score_func=auc_score)\end{lstlisting}
\end{lrbox}%
\scalebox{0.92}{\usebox{\mybox}}

\begin{mytbox2}[colback=green!15,fontupper=\scriptsize,flushleft upper,boxrule=0pt,arc=5pt,left=2pt,right=2pt,after=\ignorespacesafterend\par\noindent]{\checkmark Accepted Answer}
	As noted already, for \texttt{SVM}-based Classifiers (as \texttt{y == np.int*}) \textbf{preprocessin}g is a must, otherwise the ML-Estimator's prediction capability is lost right by skewed features' influence onto a decission [\emph{sic}] function.
\end{mytbox2}
\captionof{Stack Overflow post}{\href{https://stackoverflow.com/questions/17455302/gridsearchcv-extremely-slow-on-small-dataset-in-scikit-learn}{Example post with contract} \label{fig:post4}}
\hspace*{0.3cm}

\begin{description}[leftmargin=*]
\item [\bfseries SAM-AMO Inter-dependency (SAI):] SAI contracts have a dependency between behavior and method orders. This dependency could be in either direction, i.e., the program's state could determine the order of API calls, or the order of API calls could require that some condition must hold. For example, in \SO post~\ref{fig:post4}, if an \ML API users uses the SVM-based classifier \texttt{SVC} as the \emph{estimator} parameter for \texttt{GridSearchCV()} with \scikit, then \texttt{preprocessing.scale()} must precede this call. Since the order of the method calls \texttt{GridSearchCV()} and \texttt{preprocessing.scale()} APIs is dependent upon the value given to the parameter of \texttt{GridSearchCV()}, it belongs to the \emph{SAI} contract category.
The root cause of this contract violation is that the value being passed to one method call (\texttt{GridSearchCV()}) requires a temporal ordering between the two (\texttt{GridSearchCV()} and \texttt{preprocessing.scale()}) methods. 
\end{description}

The leaf components of this subclass contain all contract cases that we derived individually for the \emph{SAM and AMO} categories. For example, if an \emph{intra-argument contract, IC-1} of an API determines an \emph{always (G),} order of two APIs like the example above then, it belongs to \textbf{SAM (Level 3) $\land$ AMO (Level 2)}. 

Any dependency between SAM related leaf nodes, e.g., primitive type, built-in type, \ML~type, etc. and AMO related leaf nodes, i.e., G and F, belong to this category. 

\begin{description}[leftmargin=*]
	\item [\bfseries Selection (SL):] The final subclasses in our classification are those contracts that involve a choice when it comes to enforcing an API related contract. If the choices only belong to the contract components of \emph{SAM} or \emph{AMO}, then we categorize the contracts into either \textbf{SAM (Level 3)} or \textbf{AMO (Level 2)}, respectively. For instance, in \SO post \ref{fig:postSL1}, 
	the author wants to convert two \texttt{numpy} arrays to \texttt{Tensor}s and uses \texttt{TensorDataset()} from the \torch library. The arguments of this API must either be of \texttt{Double} or \texttt{Float} type. The API then confirms \texttt{DoubleTensor} conversion upon exiting. Hence, there are two choices of category SAM (specifically IC-2) to maintain the contract for this API, and we mark this with \textbf{SAM (Level 3) category}. The root cause of this contract violation is that the client did not follow one of the two choices (providing arguments of \texttt{Double} or \texttt{Float} types). \\
	\hspace*{0.2cm}
	\begin{lrbox}{\mybox}%
\begin{lstlisting}[language=iPython,numbers=none]
train = data_utils.TensorDataset(torch.from_numpy(X).double(), torch.from_numpy(Y))
train_loader = data_utils.DataLoader(train, batch_size=50, shuffle=True)
for batch_idx, (data, target) in enumerate(train_loader):
    data, target = Variable(data), Variable(target)
    optimizer.zero_grad()
    output = model(data)               # error occurs here
\end{lstlisting}
\end{lrbox}%
\scalebox{0.92}{\usebox{\mybox}}

\begin{mytbox2}[colback=green!15,fontupper=\scriptsize,flushleft upper,boxrule=0pt,arc=5pt,left=2pt,right=2pt,after=\ignorespacesafterend\par\noindent]{\checkmark Accepted Answer}
The \texttt{numpy} arrays are \texttt{64-bit} floating point and will be converted to torch. \texttt{DoubleTensor} standardly. Now, if we use them with our model, we'll need to make sure that your model parameters are also Double. Or we need to make sure, that your \texttt{numpy} arrays are cast as \texttt{Float}, because model parameters are standardly cast as \texttt{float}.
\end{mytbox2}
\captionof{Stack Overflow post}{\href{https://stackoverflow.com/questions/44717100/pytorch-convert-floattensor-into-doubletensor} {Example post with contract} \label{fig:postSL1}}
\hspace*{0.3cm}
	
	Another example can be seen in the \SO post \ref{fig:postSL2},  
	where the author of the post is using the \keras library to create a neural network. Then they want to initialize and fit the neural network weights and save these weights. Next, they want to use these saved weights and predict some output values given the inputs. However, they had issues using the \texttt{load\_weights()} API to collect the saved weights. The answer post explains that as one uses the \texttt{load\_weights()} API, one has to maintain an order between two other related APIs (\texttt{compile} and \texttt{predict}). One expected order is calling \texttt{load\_weights()}, \texttt{compile()}, \texttt{predict()}. The order alternative is calling \texttt{compile()}, \texttt{load\_weights()}, and \texttt{predict()} at some point in history. As both choices involve AMO, this belongs to the \textbf{AMO (Level 2)} category. The root cause of this contract violation is that the client did not make one of the two choices (maintaining the method order between the related APIs). \\
 %\newpage
	\hspace*{0.2cm}
	\begin{lrbox}{\mybox}%
\begin{lstlisting}[language=iPython,numbers=none]
model = Sequential()
...
model.load_weights('keras_w') 
y_pred = model.predict(X_nn)
\end{lstlisting}
\end{lrbox}%
\scalebox{0.92}{\usebox{\mybox}}

\begin{mytbox2}[colback=green!15,fontupper=\scriptsize,flushleft upper,boxrule=0pt,arc=5pt,left=2pt,right=2pt,after=\ignorespacesafterend\par\noindent]{\checkmark Accepted Answer}
We need to call \texttt{model.compile}. This can be done either before or after the \texttt{model.load\_weights} call but must be after the model architecture is specified and before the \texttt{model.predict} call.
\end{mytbox2}
\captionof{Stack Overflow post}{\href{https://stackoverflow.com/questions/33474424/keras-load-weights-of-a-neural-network-error-when-predicting} {Example post with contract} \label{fig:postSL2}}
\hspace*{0.3cm}

In comparison, if the choices involve both SAM and AMO, then we categorize the contract as a combination type contract \textbf{Comb. of SAM and AMO}. For example, in \SO~post \ref{fig:post1}, we observe such a combination. The accepted answer respondent mentions two alternative ways to maintain correctness when using the \texttt{tf.random\_shuffle} API. The first choice is setting the argument \texttt{seed} for this API to some desired value. The second is maintaining an order between invocations of \texttt{tf.random\_shuffle()} and \texttt{tf.reset\_default\_graph()}. Since the same contract breach can be resolved through either a behavioral or temporal contract that involves \texttt{tf.random\_shuffle()} API, there is a selection involved as to which one to be adopted. 
	Documentation should include all choices to maintain contracts for an API method. The root cause of this contract violation is that the client did not make one of the two choices (providing an acceptable \texttt{seed} value or using an acceptable method ordering) for the API to function properly.
	Researchers should emphasize the need to be able to express such requirements to users, who can choose to satisfy the requirements of a library either by maintaining a temporal order or by some state-based change. Therefore, the practitioners can design and develop a contract checking mechanism for ML API calling orders to facilitate the end-users.
	
\end{description}
\subsection{Classification of ML Contract Violation Locations}
\label{subsec:vl}

As we investigated the requirements for \ML~contracts, we also classified the 
API locations of the contracts being violated. 

We based this classification on prior works \cite{stages, islam19}, and used a similar open coding strategy as we did when conceiving the classification for contract types. 
The categories are explained in Table~\ref{tab:class2}.

\begin{table}[h!t]
\caption{Contract Violation Locations}
\label{tab:class2}
\setlength{\tabcolsep}{3.5pt}
\scriptsize
\resizebox{\columnwidth}{!}{%
\begin{tabular}{|| p{1in}|| p{2.65in} ||}
	\hline
	\rowcolor{white}
	{\cellcolor[gray]{.9}Data Preprocessing} & These APIs pre-process data before feeding it to \ML~models. \\
	\hline
	{\cellcolor[gray]{.9}Model Construction} & These APIs are used to build \ML~models, either from scratch, by accessing a predefined model, or by compiling constructed models. \\
	\hline
	\rowcolor{white}
	{\cellcolor[gray]{.9}Model Evaluation} & APIs used to estimate the generalization accuracy of a model. \\
	\hline		
	{\cellcolor[gray]{.9}Model Initialization} & APIs used for initializing a predefined model, e.g., an API to load random weights for a model.\\
	\hline
	\rowcolor{white}
	{\cellcolor[gray]{.9}Train} & Describes APIs that determine values for weights and biases of a model.\\
	\hline
	{\cellcolor[gray]{.9}Prediction} & Describes APIs that predict an outcome after training. \\
	\hline
	\rowcolor{white}
	{\cellcolor[gray]{.9}Hyper-parameter Tuning} $\;$ & Describes APIs that change hyper-parameter(s) that control the learning process.\\
	\hline
	{\cellcolor[gray]{.9}Load} & Describes APIs that load or store data from external storage. \\
	\hline
\end{tabular}
}
\end{table}

\subsection{Classification of Effects}
\label{subsec:class}

We used a prior work \cite{islam19} for classifying the effects to the 
root causes discussed in \S{\ref{subsec:con}}. It has six categories: bad performance (BP), crash (C), data 
corruption (DC), hang (H), incorrect functionality (IF), and 
memory out of bound (MOB). We have added one category Unknown (U)
besides these categories to identify cases that remain non-classified. 
The details about this classification of effect are discussed in Table~\ref{tab:classeffect}.

\begin{table}[h!t]
%\resizebox{\textwidth}{!}{%
\caption{Contract Violation Effects}
\label{tab:classeffect}
\setlength{\tabcolsep}{3.5pt}
\begin{tabular}
{||
>{\columncolor[HTML]{EFEFEF}}l ||p{3.15in}||}
\hline
Bad Performance &
  Common effect in ML software; \ML API users face model problems even though they use deep learning APIs correctly because APIs in these libraries are abstract. \\ \hline
Crash &
  Frequent effect in ML. In fact, any kind of contract violation can lead to a Crash. A symptom of the crash is that the software stops running and prints out an error message. \\ \hline
Data Corruption &
  This happens when the data is corrupted while flowing through the network, and a user gets unexpected output. This effect is a consequence of misunderstanding the ML APIs. \\ \hline
Hang &
  Hang is caused when ML software ceases to respond to inputs due to slow hardware or inappropriate ML algorithm. Software running for a long period of time without providing the desired output is considered as a symptom. \\ \hline
Incorrect Functionality &
  It occurs when the software behaves unexpectedly without any runtime or compile-time error due to the incorrect output format, model layers not working desirably, etc. \\ \hline
Memory Out of Bound &
  ML software often halts due to the unavailability of the memory resources for the wrong model structure or not having enough computing resources to train a model. \\ \hline
Unknown &
  Sometimes the effect of ML contract violation is unknown. We have added this category for cases that remain non-classified. \\ \hline
\end{tabular}%
%}
\end{table}
\subsection{Labeling}
\label{subsec:label}

The classification schemes described in \S{\ref{subsec:con}}, \S{\ref{subsec:vl}, and \S{\ref{subsec:class}} 
were used to label all 1565 collected \SO~posts. 
First, the second and the third authors with strong \ML~background, have familiarized themselves with contract literature. The authors have all studied key papers in the area of software specification and design-by-contract methods. Then we trained these two authors to understand the classification schema with the help of some example posts. In this training process, the two authors were shown multiple examples for each category in the classification schema. The examples were demonstrative of where the contract is broken for an \ML~API and how the accepted answer describes the correct usage for that precise API.
Then, each rater performed independent labeling of these posts in two iterative rounds. The 10\% sampled data analyzed for the classification coding scheme and the first iterative round of labeling served as part of the training process for the labelers.
To measure the inter-rater agreement, we have used Cohen's Kappa 
coefficient~\cite{viera2005understanding} as labeling progressed at 
1\%, 2\%, 5\%, 10\%, and continued in this fashion. 
We have followed the methodology used in prior 
works~\cite{10.1145/1062455.1062539, CHATTERJEE2020110454, islam19, islam20repairing}
to reconcile inter-rater 
disagreements at fixed intervals. 
During first iterative round, at 5\% and 10\%, we report the Kappa coefficient to be 40\% 
and 51\%, respectively. 
The low value
of the agreement directed the raters to meet more frequently (at each 2\%) for a second iterative round during the first few intervals to clarify the labels that raters were using for each post. During these meetings,
raters discussed the reasoning behind cases where a strong disagreement occurred in a moderator's presence. We continuously checked the Kappa coefficient at these intervals, 
and 
even if the Kappa value fluctuated we 
reached values over 80\% after
completing labeling 22\% of the posts for the entire dataset. 
According to ~\cite{10.1093/ptj/85.3.257}, a Kappa coefficient 
value higher than 0.80 is considered as almost perfect agreement.

\section{Results}
\label{sec:result}

The main question we asked is about what contracts are most needed by \ML API users. In this section, we present the quantitative data (e.g., representing contract violation patterns, root cause, effect, contract comprehension challenges, etc.) to show the places where immediate support for contracts is needed. Hence, we analyze the results from our \SO~study 
to answer the research questions from \S{\ref*{sec:intro}}, 
report our findings on the original (and the filtered subset) dataset described in \S{\ref*{sec:data}}, and discuss implications and actionable insights.
\subsection{Contract Frequency, Root Cause, and Effect of Contract Violations}
\label{subsec:ctvs}

In this subsection, we answer \RQ{1} by presenting the required types of \ML contracts, the root causes of contract violations, and related effects. 

\subsubsection{\textbf{Required ML contracts and associated root causes}}

To explore required ML contracts and the root causes behind contract violations, we use the leaf contract types from our classification (\S{\ref{subsec:con}}) schema. %The statistical result in 
Table~\ref{tab:socontract1} shows the frequency of each type of contract from the classification found in our dataset. Figure~\ref{level3pie} demonstrates the corresponding root causes. Figure~\ref{fig:compare} shows the statistical comparison of ML Contracts for two datasets (all filtered posts and the subset containing posts with scores of 30 or higher).

\finding{Most frequent ML API contracts are: \newline
	\textbullet~constraint check on single arguments of an API. \newline
	\textbullet~order of APIs that become a requirement eventually. 
}

\begin{table}[htbp!]
	\centering
	\caption{Statistics of \ML~Contracts in \SO}
	\label{tab:socontract1}
	\footnotesize
\resizebox{\textwidth}{!}{%
\begin{tabular}{|l|rrrr|r|}
\hline
\rowcolor[HTML]{F3F3F3} 
\cellcolor[HTML]{EFEFEF} &
  \multicolumn{4}{c|}{\cellcolor[HTML]{F3F3F3}\textbf{\bfseries ML Library}} &
  \multicolumn{1}{l|}{\cellcolor[HTML]{EFEFEF}} \\ \cline{2-5}
\rowcolor[HTML]{F3F3F3} 
\multirow{-2}{*}{\cellcolor[HTML]{EFEFEF}\textbf{Contract Types}} &
  \multicolumn{1}{r|}{\cellcolor[HTML]{F3F3F3}\tf} &
  \multicolumn{1}{r|}{\cellcolor[HTML]{F3F3F3}\scikit} &
  \multicolumn{1}{r|}{\cellcolor[HTML]{F3F3F3}\keras} &
  \torch &
  \multicolumn{1}{l|}{\multirow{-2}{*}{\cellcolor[HTML]{EFEFEF}\begin{tabular}[c]{@{}l@{}}Overall \end{tabular}}} \\ \hline
Primitive Type (PT) &
  \multicolumn{1}{r|}{0.63\%} &
  \multicolumn{1}{r|}{1.65\%} &
  \multicolumn{1}{r|}{0.97\%} &
  0.00\% &
  0.01\% \\ \hline
\rowcolor[HTML]{EFEFEF} 
Built-in Type (BIT) &
  \multicolumn{1}{r|}{\cellcolor[HTML]{EFEFEF}1.88\%} &
  \multicolumn{1}{r|}{\cellcolor[HTML]{EFEFEF}5.79\%} &
  \multicolumn{1}{r|}{\cellcolor[HTML]{EFEFEF}1.94\%} &
  3.85\% &
  3.18\% \\ \hline
Reference Type (RT) &
  \multicolumn{1}{r|}{0.63\%} &
  \multicolumn{1}{r|}{2.48\%} &
  \multicolumn{1}{r|}{3.88\%} &
  3.85\% &
  2.20\% \\ \hline
\rowcolor[HTML]{EFEFEF} 
ML Type (MT) &
  \multicolumn{1}{r|}{\cellcolor[HTML]{EFEFEF}15.00\%} &
  \multicolumn{1}{r|}{\cellcolor[HTML]{EFEFEF}14.05\%} &
  \multicolumn{1}{r|}{\cellcolor[HTML]{EFEFEF}16.50\%} &
  15.38\% &
  15.16\% \\ \hline
Intra-argument Contract (IC-1) &
  \multicolumn{1}{r|}{20.63\%} &
  \multicolumn{1}{r|}{33.88\%} &
  \multicolumn{1}{r|}{34.95\%} &
  23.08\% &
  28.36\% \\ \hline
\rowcolor[HTML]{EFEFEF} 
Inter-argument Contract (IC-2) &
  \multicolumn{1}{r|}{\cellcolor[HTML]{EFEFEF}3.75\%} &
  \multicolumn{1}{r|}{\cellcolor[HTML]{EFEFEF}1.65\%} &
  \multicolumn{1}{r|}{\cellcolor[HTML]{EFEFEF}0.97\%} &
  3.85\% &
  2.44\% \\ \hline
Always (G) &
  \multicolumn{1}{r|}{11.25\%} &
  \multicolumn{1}{r|}{7.44\%} &
  \multicolumn{1}{r|}{7.77\%} &
  11.54\% &
  9.29\% \\ \hline
\rowcolor[HTML]{EFEFEF} 
Eventually (F) &
  \multicolumn{1}{r|}{\cellcolor[HTML]{EFEFEF}19.38\%} &
  \multicolumn{1}{r|}{\cellcolor[HTML]{EFEFEF}15.70\%} &
  \multicolumn{1}{r|}{\cellcolor[HTML]{EFEFEF}10.68\%} &
  23.08\% &
  16.38\% \\ \hline
SAM (Level 3) $\land$ AMO (Level 2) &
  \multicolumn{1}{r|}{7.50\%} &
  \multicolumn{1}{r|}{8.26\%} &
  \multicolumn{1}{r|}{7.77\%} &
  0.00\% &
  7.33\% \\ \hline
\rowcolor[HTML]{EFEFEF} 
SAM (Level 3) &
  \multicolumn{1}{r|}{\cellcolor[HTML]{EFEFEF}4.38\%} &
  \multicolumn{1}{r|}{\cellcolor[HTML]{EFEFEF}2.48\%} &
  \multicolumn{1}{r|}{\cellcolor[HTML]{EFEFEF}0.97\%} &
  3.85\% &
  2.93\% \\ \hline
AMO (Level 2) &
  \multicolumn{1}{r|}{6.25\%} &
  \multicolumn{1}{r|}{1.65\%} &
  \multicolumn{1}{r|}{5.83\%} &
  3.85\% &
  4.65\% \\ \hline
\rowcolor[HTML]{EFEFEF} 
Comb. of SAM (Level 3) and AMO (Level 2) &
  \multicolumn{1}{r|}{\cellcolor[HTML]{EFEFEF}8.75\%} &
  \multicolumn{1}{r|}{\cellcolor[HTML]{EFEFEF}4.13\%} &
  \multicolumn{1}{r|}{\cellcolor[HTML]{EFEFEF}7.77\%} &
  7.69\% &
  7.09\% \\ \hline
\end{tabular}%
}
\end{table}

\begin{figure}[h]
    \includegraphics[width=4.5in,trim={0cm 0cm 0cm 0cm},clip]{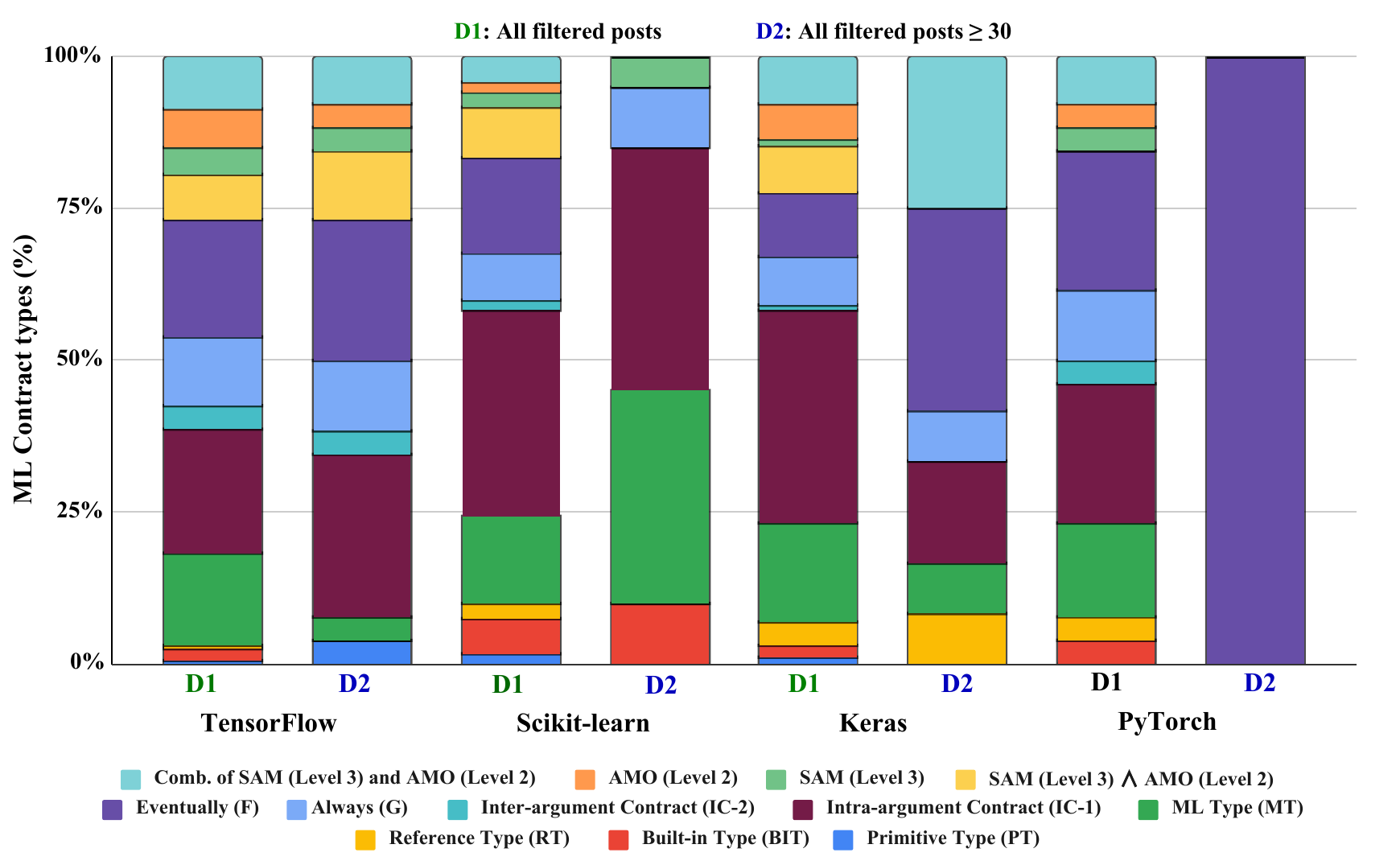}
    \caption{Comparison of \ML Contract types of all filtered posts (D1) and subset with score $\ge$ 30 (D2) in \ML libraries}   
    \label{fig:compare}
\end{figure}
\phantomsection
\label{par:IC-1}
{\textbf{Required \ML~Contract.}} We identify that breaking the contract on the \emph{single argument of an API (IC-1)} and \emph{eventually (F) required API method orders} are the most frequent type of contracts violated. 
We observe that the lack of domain knowledge, and incomplete error messages are some of the reasons why \ML API users struggle with the IC-1 category. For example, in \SO~post \ref{fig:post1} the author struggled to grasp the difference between graph level seed and operation level seed when using the \texttt{tf.random\_shuffle} API. 
In addition, some \ML~APIs are involved in \emph{AMO} contracts that require particular method orders. This required method order is often a source of confusion. For the posts with score $\ge$ 30 in \torch library (\ref{fig:compare}), all observed contracts belong to \emph{AMO} category. However, the number of posts with a score of 30 and higher and containing a contract from the \torch library is very low (3 contracts). Thus, we refrain from making any additional observations for this case.
To analyze further why the required contracts mentioned in this finding are commonly violated, we have randomly sampled \ML APIs from our dataset and studied the documentation for these APIs to investigate if the documentation is complete. We have analyzed API documentation from the \keras~and \tf~libraries and observed that 
many of these incorrect usages of APIs are not documented, especially the corner cases.
As an instance, the function \texttt{RELU} is a valid activation function for \ML~layer APIs in \tf.
However, it should not be used if the layer API in question is the output layer of the 
model in a multi-label classification. 

The SE community can employ existing contract mining approaches~\cite{parammine, specmining, ltlmining, lemieux2015mining, deepspecs} 
to mine these contracts and enhance library documentation.

\begin{figure}[h]
	\includegraphics[width=4.5in,trim={0.5cm 0.5cm 0.5cm 0cm},clip]{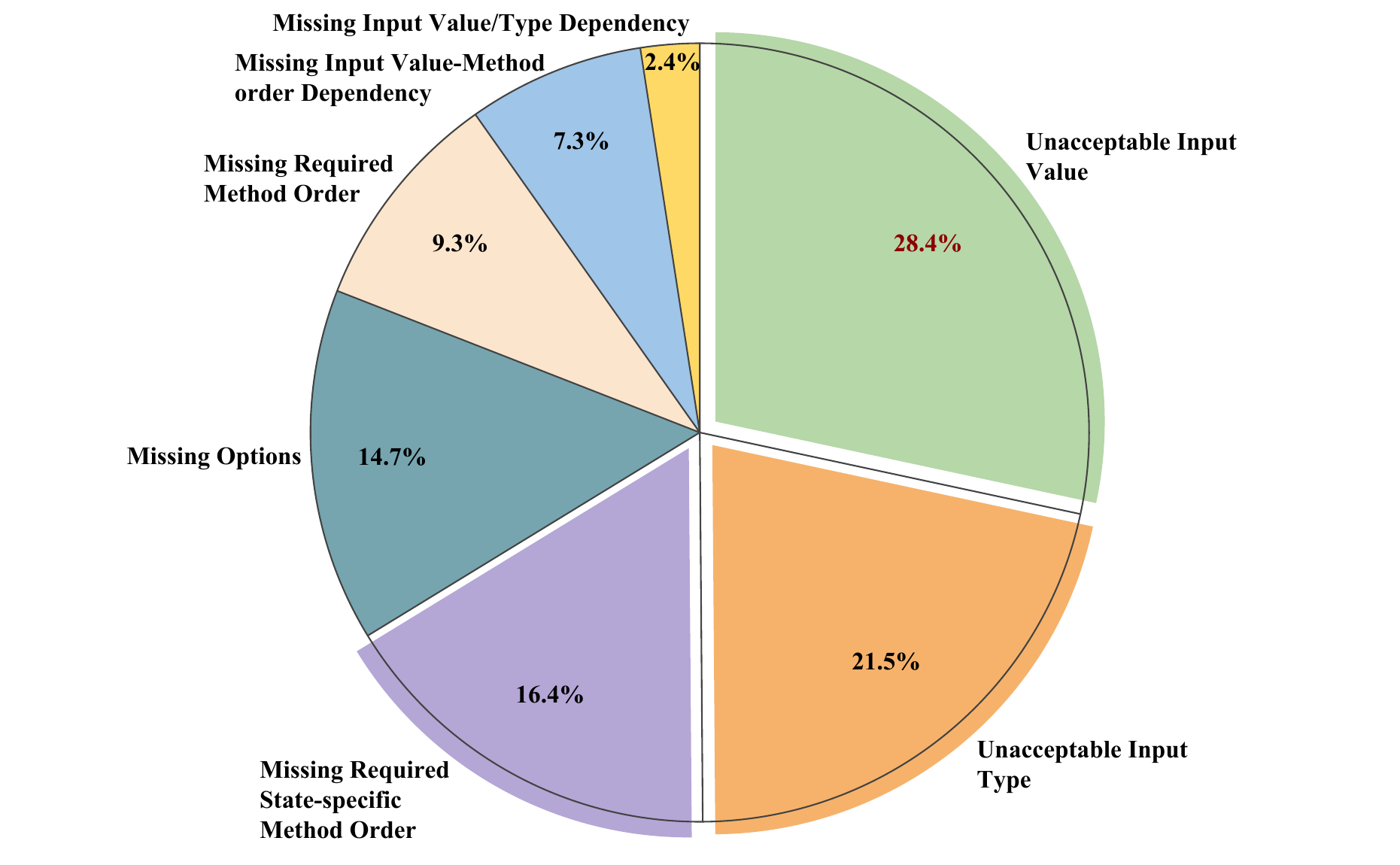}
	\caption{Distribution of root causes behind \ML~contract violations}
	\label{level3pie}
\end{figure}

\finding{\ML~APIs require \ML type checking contracts and show inter-dependency between behavioral and temporal contracts.}
%\vspace{0.05cm}
%\noindent
\phantomsection
\label{par:ML}

We have observed ML type checking (MT) is the next major category, 
considering all posts. For instance, in one \SO~post\footnote{https://stackoverflow.com/questions/33762831/}, the \ML API users is trying to use a predefined model through a \tf~API \texttt{seq2seq()}. This API essentially consists of two recurrent neural networks. The encoder part processes the input and the decoder generates the output. To capture this, \texttt{seq2seq()} contains two arguments \texttt{encode\_inputs} and \texttt{decode\_inputs}. The contract requirement for these arguments according to the accepted answer is that if the input has some shape \texttt{[n]}, then both of the arguments are required to have a shape of \texttt{[batch\_size $\times$ n]}.
We also note that, the ML-type checking error is more common (for the posts with 
score $\geq30$) in the \scikit~library compared to other libraries. This is one of the key findings that is different when we compared the original curated dataset and the filtered dataset with posts scored 30 or higher. This observation can be attributed to the fact that the other studied \ML~libraries incorporate some type checking system, unlike~\scikit. As a result, a \tf or a \keras or a \torch program is less likely to contain type errors. For instance,  we have described that in the \SO post~\ref{fig:post2}, the \texttt{matmul()} API from the \tf library requires that both of the arguments assume that the same \texttt{Tensor} types will be provided by the caller of the API. Therefore, supplying anything other than the allowed type will cause a type error and the program to crash. In contrast, \scikit does not require its program to be strongly typed and relies on Python’s default type system. This situation highlights the need for type regulation in the \ML framework. A runtime assertion checking tool could help catch such contract violations; such a tool could be built, for example, by enhancing an off-the-shelf (e.g., PyContracts~\cite{pycontracturl}) tool that can detect violations of the ML-type contracts we propose. We note that some of the type issues may be caused by the dynamically typed nature of the programming language, Python, and are out of the scope for this paper. 

Additionally, we see that one other new category (dependency between behavioral and temporal contracts) in our classification is required for significant number of APIs. 
Contract languages and type checking tools~\cite{pythonstaticchecker, seshia2018formal, jothimurugan2019composable} should add sufficient expressiveness for these additional types of contracts seen in \ML~APIs.

We note that the behavioral contracts reported in this study are largely preconditions. However, we found some postconditions as well. For instance, in SO post\footnote{https://stackoverflow.com/questions/40904979/}, the author is using the \texttt{tensorflow.session()} method to return a \texttt{Session} object. A \texttt{Session} is a class that is used to run \tf operations. Then calling the \texttt{run()} method on this \texttt{Session} object allows evaluation of the \texttt{Tensor}. In the example post cited above, the \texttt{Tensor} is a constant \texttt{String}. It is supplied as the value used for the argument \texttt{Fetches}. We know that any value in a \texttt{Tensor} holds the same data type with a known (or partially known) shape. In this example, the value returned by \texttt{run()} has the same shape as the \texttt{Fetches} argument. Now one can decode this output data as needed.  The contract we see in this~\SO post is a postcondition. The contract for the \texttt{tf.io.decode\_raw()} API is "returns a binary string (python 2), byte string (python 3)" upon exiting the API call. \\

\finding{Unacceptable input value is the most common root cause.}
%\vspace{0.05cm}
\textbf{Primary Root Cause.} We identify that supplying unacceptable input values to APIs is the primary root cause behind contract violation in \ML. The \ML API users fail to recognize acceptable input values often for several reasons, e.g., misunderstanding a hyper-parameter setting. The undesired input values found in our study can be utilized as test cases in \ML~systems and avoid some of these contract breaches. 

\subsubsection{\textbf{Effects of Contract Violations}} To realize the effects of the contract violations, we have used the classification of effects from a prior work \cite{islam19} mentioned in \S{\ref{subsec:class}}. Figure~\ref{fig:effect} illustrates the distribution of contract violation effects across libraries.

\begin{figure}[h]
    \includegraphics[width=4.6in,trim={0.1cm 0.1cm 1.1cm 1.5cm},clip]{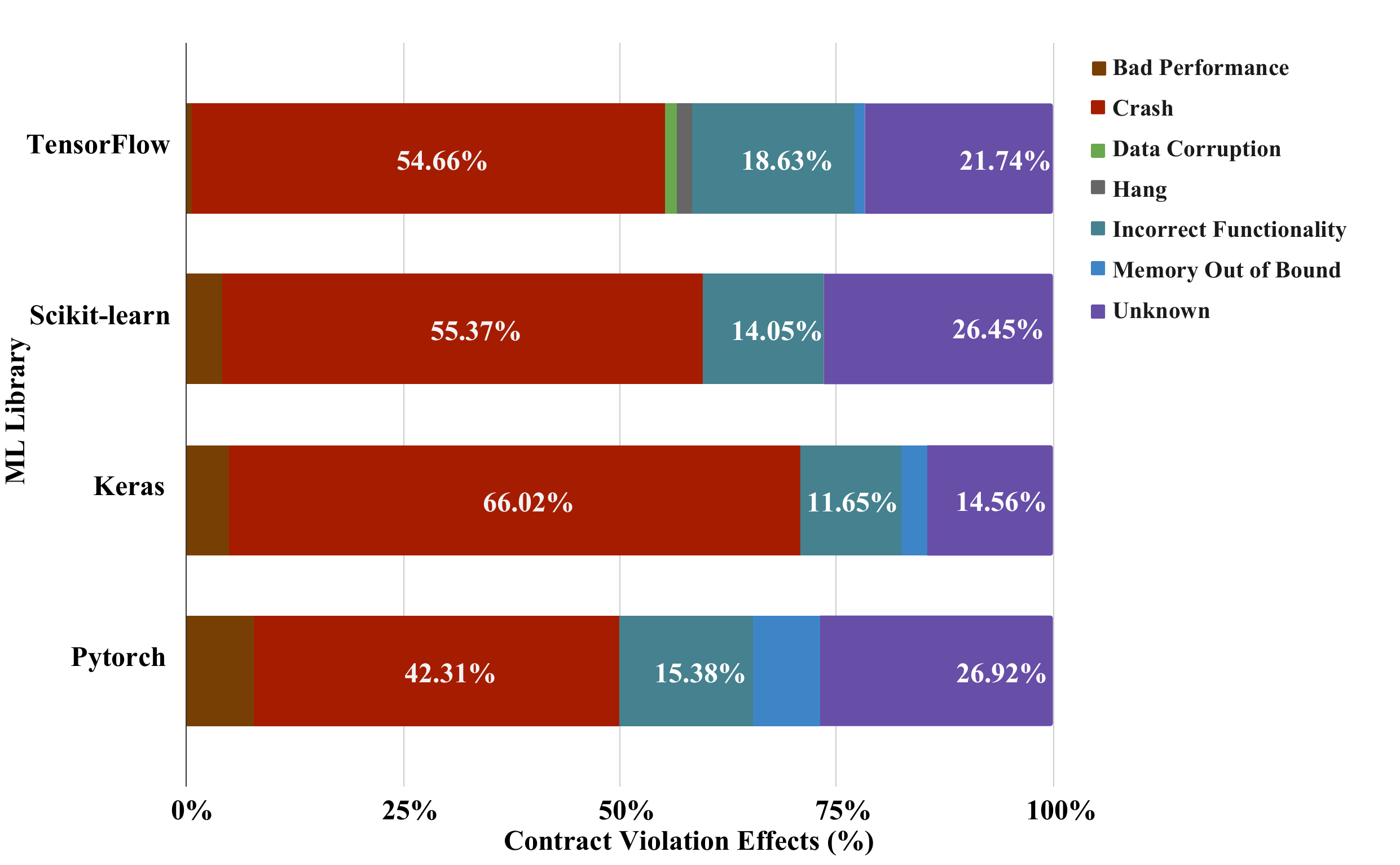}
    \caption{Distribution of \ML~contract violation effects} 
    \label{fig:effect}
\end{figure}

\finding{On average, 56.93\% of the contract violations for the \ML~libraries leads to a crash.}

{\textbf{Crash (C).}} The majority of contract violations for \ML~APIs lead to a crash in the software and we observe the range within 42.31\%-66.02\%. This result varies only for \torch~ with post score $\geq30$. 
\scikit~has the most examples of contract violations that have lead to crashes among the chosen libraries. As an instance, in \SO post~\ref{fig:post2}, violating the inter-argument contract (IC-2), would result in a crash for the program. 
Researchers might build a new automated repair tool inspired by existing repair tools ~\cite{genprog, angelix, 6776507, spr}. In this regard, mined contracts could be utilized that lead to crashes and act as a preemptive measure for the code. 

\finding{Incorrect functionality has a different frequency pattern in \keras.}
\noindent
{\textbf{Incorrect Functionality (IF).}} 
The next frequent (15.33\% on average) effect that we observed in terms of contract violation is causing incorrect functionality in \ML~software. \ML API users apprehend the deviant behavior either from experience or through the logical organization of the \ML~model.

However, the library \keras~shows a lower percentage. This divergence in frequency can be explained by the fact that \keras~is a high-level deep learning library; thereby hiding many complicated implementation details 
which reduces the chance of running into IF. Since the compiler cannot catch incorrect functionality, our dataset can become a benchmark through contract annotation to detect this effect. Expert \ML API users can further rank these particular contract violations (e.g., scary, troubling, concern) to provide a hint of severity for these IFs as many bug detection tools (such as Bugram \cite{bugram}, Salento \cite{salento}) do. 

\finding{Contract violation-effect  distribution is similar across libraries.}
\noindent
{\textbf{Contract violation-effect correlation.}} To determine the effect of breaking a contract, we used the information explicitly available in the \SO~post we labeled. Even so, we hypothesize that by observing the type of contract violation, it is possible to make an informed guess on the corresponding effect. 
For instance, Islam \etal reported that the violation of type or shape usually results in a crash during runtime~\cite{islam19}.
One \SO post~\footnote{https://stackoverflow.com/questions/44429199/} author was not sure how to use the API \texttt{DataLoader()} from the \torch~library. The answer post lists that the API in question requires that the argument type should be a subclass of \texttt{Dataset} class. Even though it was not explicitly described on the post, such a type checking contract violation would result in a crash. \textit{Python} being a loosely typed language, type mismatch may go unnoticed during compilation. it usually crashes for mismatches in the expected type or shapes~\cite{islam19}.
	
To test this hypothesis on learning from other \ML~libraries regarding violation-effect correlation, 
we obtained the conditional probabilities, %\\ 
%\hspace*{3.5 cm}
\begin{equation*}
\Pr(E=\mathrm{effect}_\mathrm{i} \mid V=\mathrm{violation}_\mathrm{j})
\end{equation*}
\\ which describe how likely a certain effect ($\mathrm{effect}_\mathrm{i}$) follow given a contract violation ($\mathrm{violation}_\mathrm{j}$). Then, we utilized the \emph{Jensen–Shannon divergence (JSD)}~\cite{endres2003new} measure to compute the distance between two probability distributions, $E$ and $V$. The divergence ranges from 0 to 1, where 0 indicates perfect similarity, and 1 indicates no similarity. 

In our experiment, we observed that the violation-wise effect distribution is similar across chosen libraries. The result shows that %among the four contract types with 10\% support, 
eventually (F) required method order, \ML type checking (MT), and intra-argument contracts (IC-1) demonstrated a divergence score of 0.08, 0.11, and 0.14, respectively, with 10\% support, indicating a similar effect distribution across libraries. 
This experiment agrees with our hypothesis.   
Therefore, the SE community can learn from contract violations of the same category for \ML~libraries and estimate unexpected behaviors of other \ML~libraries with similar effects in code. Furthermore, this experiment also shows an application of the proposed classification schema.

\finding{Error messages thrown for breaching an \ML~contract are not often adequate at present.}

%\vspace{0.2cm}
%\noindent
\phantomsection
\label{par:Error}
{\textbf{Error message.}} In case of system failure, the crash or error message helps API users debug the code and identify the root cause. 
In the \SO~post \ref{fig:post8}, the author had received the
error message in the listing below when they tried to load weight on a predefined model. It could be an exhausting task 
to understand the problem by only examining the error message. 
The answer to this post registers that the error occurs as the \ML API users missed redefining the model architecture before loading weights.  
We find an error message inadequate if the error message is present in the author's post, and the response demonstrates that the error message presented does not reflect the incorrect usage for the API in question. 
Additionally, since domain experts can explain these challenging ML contracts (see \S{\ref{sec:dcc}}), such extracted contracts can be encoded in a contract-checking tool. Such a tool, as a result, can enable improved debugging mechanisms for ML software developers.\\

{ \quad
	\begin{lrbox}{\mybox}%
		\begin{lstlisting}[language=iPython2]
IndexError    Traceback (most recent call last)
<ipython-input-101-ec968f9e95c5> in <module>()
    1 model2 = Sequential()
--> 2 model2.load_weights("/Users/Desktop/SquareSpace/weights.hdf5") ...
IndexError: list index out of range
\end{lstlisting}
	\end{lrbox}%
\scalebox{0.92}{\usebox{\mybox}}
\captionof{Stack Overflow post}{\href{https://stackoverflow.com/questions/35074549/how-to-load-a-model-from-an-hdf5-file-in-keras}{Example post demonstrating inadequate error message} \label{fig:post8}}}
\hspace*{0.2cm}

In our study, we found only a handful of contract violations that require runtime checks. For example, if overfitting happens during training, \emph{regularization-related} APIs are necessary for the \ML~model stack. Additionally, we have found cases where runtime checks against the state alone are insufficient without more context. For example, in \keras,~it is required to call \texttt{BatchNormalization()} between the \emph{linear} and \emph{non-linear} layer APIs in the model to achieve better performance. Thus, in this case, the presence of temporal history and an assertion check is required. For such kinds of \ML contracts, we can extend the traditional design-by-contract approach~\cite{10.1109/2.161279} to assert those contracts during runtime and utilize the contract violation message accordingly to inform the correct usage of \ML APIs to the users.

In summary, we observed that the majority of the \ML contracts are similar to traditional contracts, and Finding 1 indicates this. The contracts involving \ML type checking and dependency between behavioral and temporal contracts are specific and needed by ML software. Interestingly, we report there are contracts that can be formalized as in traditional contracts, however, the contract violation effect is often different, e.g., bad performance, incorrect functionality, etc.; i.e., issues about performance and accuracy are more common in \ML software.

\subsection{Common Patterns for  Contracts}

This section highlights common patterns of contracts in the dataset, i.e., we analyze the common patterns of ML contract violations observed in our study.
In section \S{\ref{subsec:ctvs}}, we noted that IC-1, F, MT are the most frequently occurring patterns across libraries. 
These contracts are \emph{atomic} in the sense that there is no dependency between behavioral and temporal contracts in these. We further investigated more complex contract patterns, including combinations of two or more atomic contracts, when answering \RQ{2}. These types of patterns belong to the high-level category \emph{hybrid} in our classification schema. Recall that hybrid contracts contain combinations, choices, or dependencies between the behavioral and temporal contracts. 

\begin{figure}[h]
    \includegraphics[width=4.7in,trim={0cm 2.4cm 0cm 1.5cm},clip]{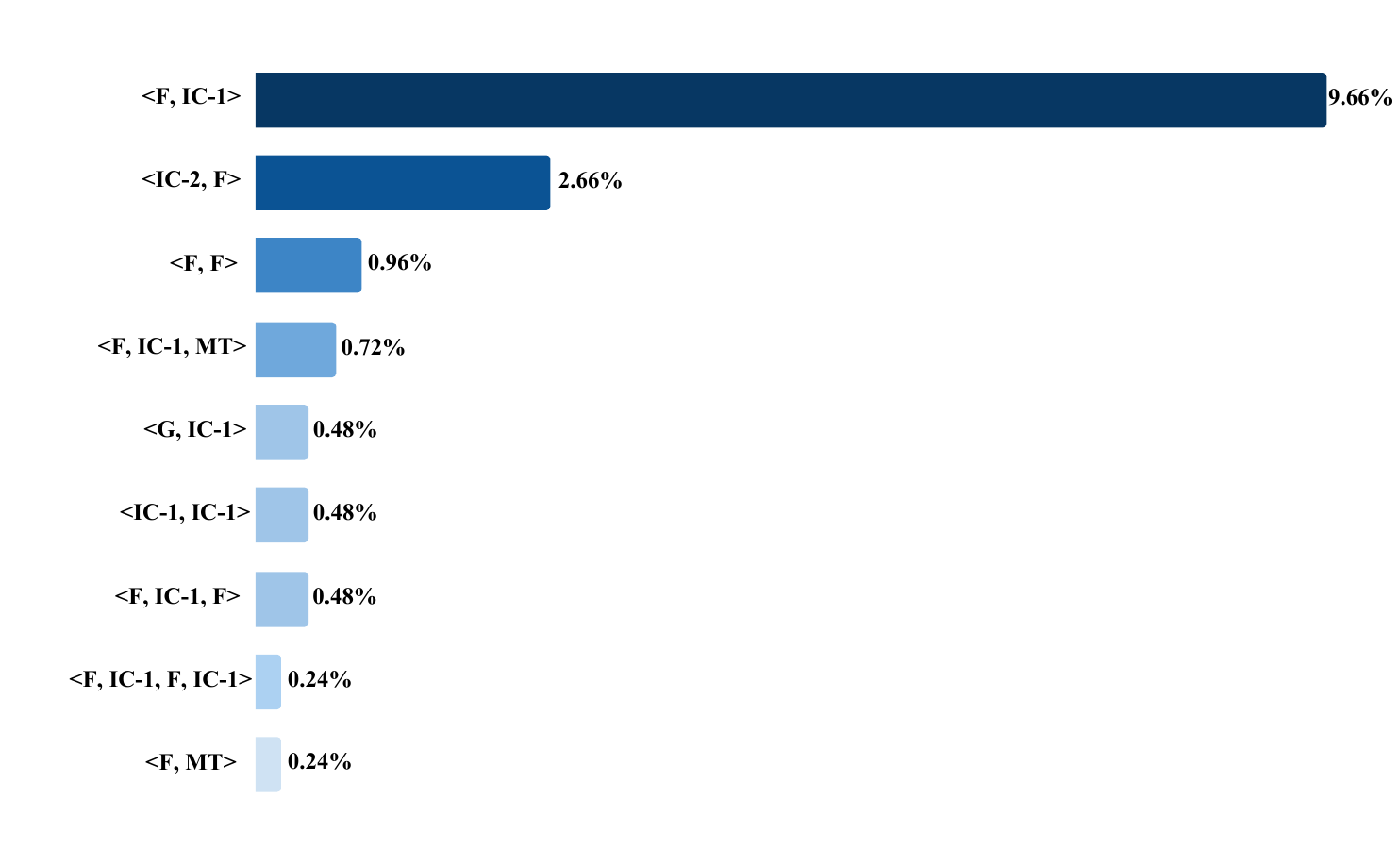}
    \caption{Patterns of \ML~Contracts}   
    \label{fig:level3pattern}
\end{figure}

\finding{Eventually (F) related hybrid contracts are one of the most common patterns across ML libraries.}

\noindent
{\textbf{Patterns involving <F>.}} Our result shows that F contracts (about the method orderings at a certain point in history) are the places \ML API users struggle most, compared to G (always orderings); see Figure~\ref{fig:level3pattern}. For instance, we described that in \SO~post \ref{fig:post4}, the parameter choice for the \texttt{GridsearchCV()} API dictates whether it must be preceded by \texttt{preprocessing.scale()} API. In contrast, it is not always obvious for patterns involving F, e.g., \ML API users sometimes use a pooling layer after a convolution layer to downsample the feature collected in the previous layer. Thus, this order is not mandatory for all program points. But, if the order is used\footnote{https://stackoverflow.com/questions/34092850/}, then the API user should make sure the parameter \texttt{strides} of \texttt{tf.nn.conv2d()} is compatible with the  \texttt{ksize} and \texttt{strides} parameters of the pooling layer (e.g., \texttt{tf.nn.max\_pool()}) in \tf.~Violations of such hybrid patterns can be found using unit tests that capture variants of these patterns and testers should be aware of capturing these variants.

\subsection{Difficulty in Contract Comprehension}
\label{sec:dcc}

In this section, we analyzed when the challenges are observed behind understanding \ML contracts. Discovering categories of \ML contracts is challenging because a significant number of \SO~posts do not contain contracts with accepted answers. Furthermore, the \SO~reputation of the user giving an answer does not necessarily determine whether an answer reveals a challenging \ML contract. To discover the correlation between \ML~contracts and user expertise in terms of the contract violation issue and to answer \RQ{3}, we have conducted an experiment that discerns respondent reliability and resolving time.  
We gather evidence from various perspectives described below to develop an educated guess. Prior work ~\cite{zhang2019empirical} analyzes the scores of \SO~post to comprehend the types of deep learning questions are more difficult. 

Inspired by that study, we have leveraged the \SO~reputation score to determine the overall expertise of a user. In our study, we slightly adapted the \SO reputation score, and named it the \emph{reliability score}. The reputation metric is often used to measure a user's expertise level on \SO, because it summarizes the overall impression of that user’s \SO activity. We observe that a user can earn a reputation for various topics unrelated to \ML-related skill sets. As a result, this metric poses a significant threat when we want to assess the expertise of a user in resolving ML contracts specifically. The \emph{Reliability score} tries to mitigate this threat to an extent.  Moreover, our adapted metric incorporates the number of accepted answers into account for higher confidence in the metric.

As an example, let us say, we are interested to know if a particular user from an \SO~post\footnote{https://stackoverflow.com/users/5098368/} is an expert in~\keras. While the user has a high reputation score \emph{(26,164)}, they only have a score of \emph{6} when filtered through the \keras~tag from answering two questions. This indicates that this user has accumulated most of his reputation from other areas. So, to measure the expertise level of a user, we consider their score only on relevant tags. Since we have only included posts having \textit{accepted} answers for this study, we refined this score to prioritize users having more accepted answers, which we call their \emph{reliability score}, measured as follows:

\begin{equation*}
\resizebox{0.9\columnwidth}{!}{
	$reliabilityScore$ $=$ $totalScore\times \frac{(totalAcceptedAnswer+\mathbb{C})}{(totalAnswer+\mathbb{C})}$.
}
\end{equation*}

Since a user may have no accepted answers, which would reduce their
reliability score to 0; we add an equal constant value, $\mathbb{C} > 0$
to both numerator and denominator of accepted answer percentage of the reliability score to prioritize among authors who do not have accepted answers. Here, we have used 1 as the value of $\mathbb{C}$ in the study. For example, suppose two users $A$ and $B$ have obtained a total score of 1200 and 80 respectively by answering an equal number of questions without any accepted answers. In this case, the reliability score would be zero, had it not been for the normalization constant, and both the authors would be rated equally. As author $A$ has achieved a significantly higher score compared to author $B$ for the same number of questions answered, adding $\mathbb{C}$ to the accepted answer fraction adds priority to the author with a higher answer score.

%% Table generated by Excel2LaTeX from sheet 'Merged Table'
%\begin{table}[htbp]
%  \vspace{16pt}
%  \centering
%  \setlength{\tabcolsep}{3.5pt}
%  \footnotesize
%  \caption{Expected reliability score of respondent to comprehend different contracts}
%  \renewcommand{\arraystretch}{0.5}
%    \begin{tabular}{|l|c|r|}
%    \hline
%    \cellcolor{Gray} \textbf{Library} & \cellcolor{Gray} \textbf{ML Contract (Level 3) } & \multicolumn{1}{l|}{\cellcolor{Gray} \textbf{Expected Reliability Score}} \\
%    \hline \hline
%    \multirow{6}[12]{*}{\textbf{\tf}} & IC-1  & 4.9 \\
%\cline{2-3}          & CMQ   & 3.66 \\
%\cline{2-3}          & MLT   & 5.79 \\
%\cline{2-3}          & IC-1,SPP & 5.98 \\
%\cline{2-3}          & SPP   & 6.67 \\
%\cline{2-3}          & AW    & 11 \\
%    \hline
%    \multirow{7}[14]{*}{\textbf{\scikit}} & IC-1  & 11.34 \\
%\cline{2-3}          & CMQ   & 88.35 \\
%\cline{2-3}          & BIT   & 8.53 \\
%\cline{2-3}          & MLT   & 11.86 \\
%\cline{2-3}          & IC-1,SPP & 5.21 \\
%\cline{2-3}          & SPP   & 6.79 \\
%\cline{2-3}          & AW    & 7.6 \\
%    \hline
%    \multirow{5}[10]{*}{\textbf{\keras}} & IC-1  & 6.23 \\
%\cline{2-3}          & MLT   & 5.34 \\
%\cline{2-3}          & IC-1,SPP & 5.35 \\
%\cline{2-3}          & SPP   & 13.74 \\
%\cline{2-3}          & AW    & 4.03 \\
%    \hline
%    \textbf{\torch} & SPP   & 12.4 \\
%    \hline
%    \end{tabular}%
%  \label{tab:reliable}%
%\end{table}%

\begin{table}[htbp]
	\footnotesize
	\centering
	\caption{Expected reliability score of respondents to comprehend different contracts}
	\setlength{\tabcolsep}{3pt}
%	\footnotesize
%	\renewcommand{\arraystretch}{0.9}
	\resizebox{\columnwidth}{!}{
	\begin{tabular}{|r|p{12.48em}|r|r|r|r|}
		\hline
		\multicolumn{1}{||p{5.15em}|}{\cellcolor{gray!25} Library} & \cellcolor{gray!25} \ML~Contract (Leaf) & \multicolumn{1}{p{4.15em}|}{\cellcolor{gray!25} Reliability Score} & \multicolumn{1}{p{6.945em}|}{\cellcolor{gray!25} Average Resolve Time (h)} & 
		\multicolumn{1}{|p{6.945em}|}{\cellcolor{gray!25} Average Elapsed Time (h)} & 
		\multicolumn{1}{p{6.55em}|}{\cellcolor{gray!25} First Answer Accepted (\%)} \\
		\hline
		\multicolumn{1}{|r|}{\multirow{5}[10]{*}{\tf}} & IC-1  & 5.26  & 265.33 & 238.62 & 79 \\
		\cline{2-6}          & MT   & 5.76  & 71.03 & 14.55 & 64 \\
		\cline{2-6}          & SAM(Level3)$\land$AMO(Level2) & 4.51  & 136.65 & 20.25 & 67 \\
		\cline{2-6}          & F   & 6.38  & 821.93 & 519.05 & 77 \\
		\cline{2-6}          & G    & 10.16 & 320.48 & 174.77 & 61 \\
		\hline
		\multicolumn{1}{|r|}{\multirow{3}[6]{*}{\scikit}} & IC-1  & 8.16  & 562.52 & 503.27 & 71 \\
		\cline{2-6}          & MT   & 10.88  & 788.50 & 252.18 & 59 \\
		\cline{2-6}          & F   & 8.15  & 47.85 & 0.28 & 71 \\
		\hline
		\multicolumn{1}{|r|}{\multirow{3}[6]{*}{\keras}} & IC-1  & 4.68  & 235.23 & 3.65 & 86 \\
		\cline{2-6}          & MT   & 5.18  & 19.88 & 11.10 & 82 \\
		\cline{2-6}          & F   & 8.20 & 1079.88 & 513.60 & 62 \\
		\hline
		\multicolumn{1}{|r|}{\multirow{3}[6]{*}{\torch}} & IC-1  & 6.60  & 84.38 & 0.00 & 100 \\
		\cline{2-6}          & MT   & 4.83  & 25.45 & 0.00 & 100 \\
		\cline{2-6}          & F   & 8.77 & 97.58 & 0.00 & 83 \\
		\hline
	\end{tabular}%
	}
	\label{tab:reliable}%
\end{table}%

The dataset includes \emph{average resolve time} for each type of contract, considering the time required to get accepted answers from the study and \emph{reliability score} for these respondents.

We fitted the dataset in a linear regression model first. However, it violated multiple assumptions such as linearity and normality assumptions of residuals of linear regression. Therefore, we choose the kernel ridge regression technique, a non-parametric (without any underlying assumptions about distributions of the dataset), and non-linear technique \cite{murphy2012machine}. We used radial basis function (RBF) as a kernel for fitting nonlinearity of the dataset and a gamma value of 0.1 chosen through trial-and-error analysis.
Since, in our dataset, we only included the accepted answers, the regression model predicts a minimum expected level of a user to solve the problem successfully. To that end, we use leaf contracts as features and the \emph{reliability scores} as a target variable. Considering that features are categorical data, we converted them into a one-hot encoded vector to feed into the model. Table \ref{tab:reliable} shows the expected \emph{reliability scores} and \emph{average resolve time} for a \SO~post respondent to comprehend different contracts for all \ML~libraries. For example, to respond to an intra-argument contract (IC-1) from the \tf~library, a respondent's expected  \emph{reliability score} is 5.26, and the \emph{average resolve time} is 265.33 hours. Additionally, \emph{reliability scores} are comparable for respondents within a library. Contract components with a support of less than 10\% are excluded from consideration. 

\finding{For \ML~libraries, F contracts require a higher level of expertise and a longer average time to resolve.}

A general observation is that \emph{F} contracts have respondents with comparatively higher reliability scores, ranging from 6.38 to 8.77, compared to other types of contracts. Consequently, the average resolve time for these ranges from 47 to 1080 hours (approximately). 
We reason that this difficulty is because F contracts are not as evident as G contracts, since F contracts must only eventually hold before the program terminates. 
From our dataset, it is possible to provide a benchmark for experts who can resolve F contract violations. This benchmark could be used by the \SO~forum, for example, to recommend new \ML-related posts to specific experts.
Furthermore, F contracts often rely on an implicit assumption; a significant research direction could be automating \ML~program repair tools such as DLFix \cite{dlfix} to resolve this contract violation. 

\finding{For Scikit-learn, \ML API users mostly struggle with comprehending \ML type checking.}

Surprisingly, we found that for \scikit, \ML API users mostly struggle with type checking contracts. The reliability score for this case is 10.88, and the average resolve time is 788.50 hours. We realize that \scikit provides off-the-shelf \ML~algorithms for supervised and unsupervised learning, whereas, the other DNN libraries we have chosen allow API users to implement these deep learning algorithms and neural networks. Therefore, deep neural network (DNN) \ML API users have some level of expertise towards \ML type checking compared to the API users who use higher-level \ML~libraries such as \scikit. Additionally, the DNN libraries in our study have typing rules to address type checking issues as discussed in \S{\ref{subsec:ctvs}}. There are contract-checking tools (e.g., PyContracts ~\cite{pycontracturl}) that can check simple non-\ML contracts. So, we recommend writing a similar extension tool that supports \texttt{scipy}, \texttt{CSR matrix} type checking, etc. \scikit~users can avoid type errors using such an extension tool. 
Additionally, such extensions can enforce these contracts through static or dynamic analysis.
To further verify our findings, we obtain two more measures: the average elapsed time between the post time of first response and the response that is accepted, and the percentage of time the first answer is accepted. We annotate this as \emph{average elapsed time}, and the \emph{first answer accepted} in Table \ref{tab:reliable}. A low rate of the first answer marked as accepted and higher elapsed time would generally indicate a difficulty in contract comprehension. We found that this additional evidence also confirms our finding that F contracts are usually harder to comprehend. We notice a relatively lower rate for accepting the first answer and higher elapsed time between a successful resolution and an initial attempt for the findings presented.
\subsection{Localizing Contract Violations to Pipeline Stages}

This section groups APIs into categories depending on the \ML~pipeline stage (described in  \S{\ref{subsec:vl}}) to explore \RQ{4}. \cite{islam19} report that even for a subclass of \ML contract violations that leads to bugs, bug localization is very challenging. This motivated us to study the stage of the APIs. Our goal was to identify the pipeline stages where contracts are frequently violated. Figure~\ref{fig:location} depicts the distribution of the locations where the \ML~API contract violation occurred.

\finding{A significant chunk of the ML contract violation occurred during data preprocessing and model construction stages.}

%\vspace{0.2cm}
%\noindent
\phantomsection
\label{par:eps}
{\textbf{Model Construction and Data Preprocessing.}} We observe that 30.1\% of contract violations occur during the model construction stage (across all \SO posts for all libraries). As an example, the \SO~post \ref{fig:post9} using \keras failed to use a \texttt{softmax} activation in the \emph{final} layer but chooses the value \texttt{categorical\_crossentropy} as the \emph{loss function} afterward. Here both the APIs involved, \texttt{Dense} and \texttt{Compile}, are from the model construction stage. In this case, the lack of contract checks results in the error propagating to the training and the prediction stages. 

{
\begin{mytbox2}[colback=green!15,fontupper=\scriptsize,flushleft upper,boxrule=0pt,arc=5pt,left=2pt,right=2pt,after=\ignorespacesafterend\par\noindent]{\checkmark Accepted Answer}
{\textbf{Your network will not work because of activation:}} with \texttt{categorical\_crossentropy} you need to have a \texttt{softmax} activation:
\end{mytbox2}
\captionof{Stack Overflow post}{\href{https://stackoverflow.com/questions/46204569/how-to-handle-variable-sized-input-in-cnn-with-keras}{Example post demonstrating contract violation in early \ML~pipeline stage} \label{fig:post9}}
}

%\vspace*{0.15cm}
For \scikit~and~\torch,~ we observe that 22.22\% and 22.73\% of errors occur respectively at the data pre-processing stage. Although it is one of the earliest stages in the \ML~pipeline, this trend is unique for these two libraries.

\begin{figure}[h]
    \includegraphics[width=4.5in,trim={1.5cm 0cm 1.5cm 0cm},clip]{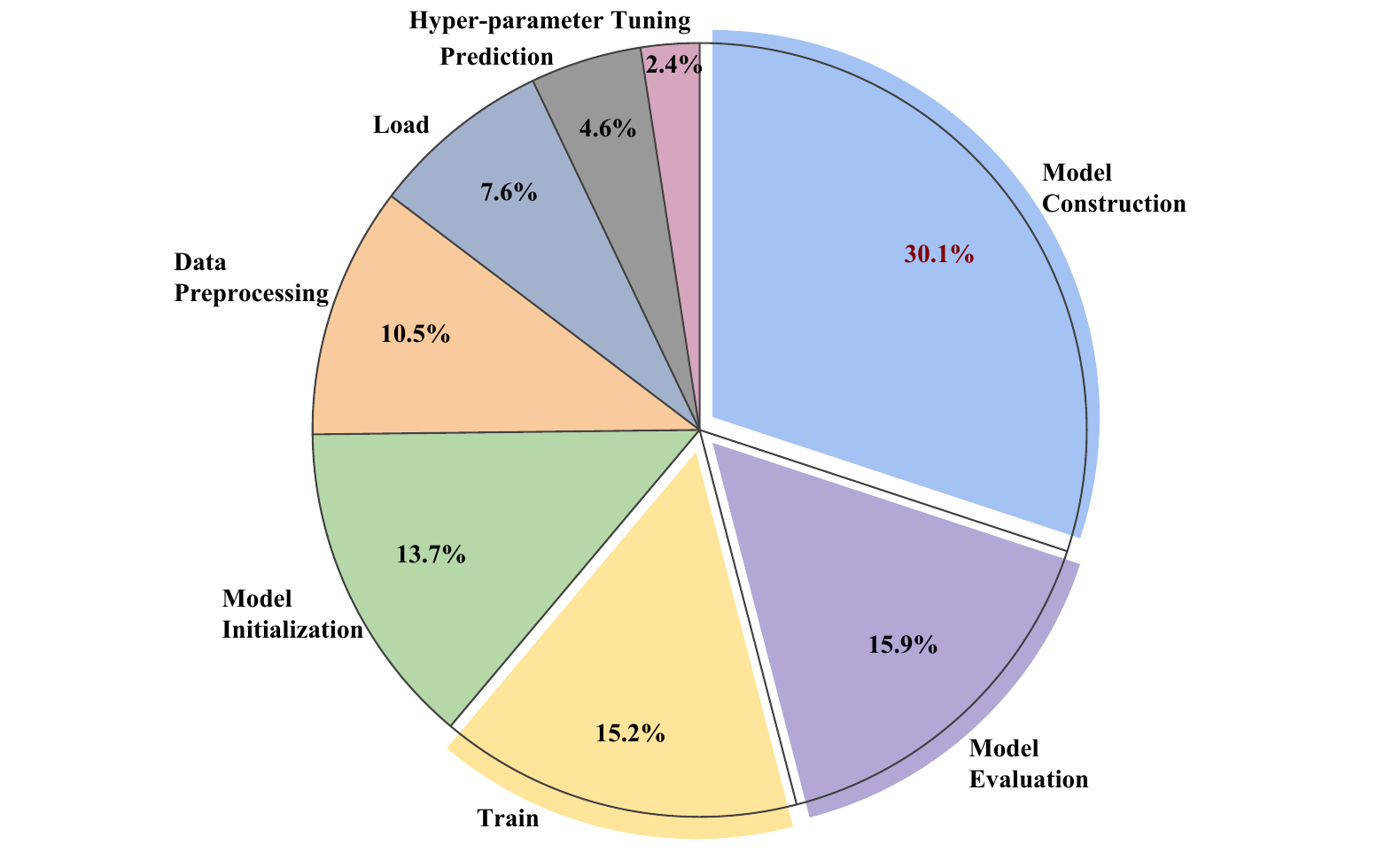}
    \caption{Distribution of \ML~contract violation stages from \SO~posts}   
    \label{fig:location}
\end{figure}

We observe that APIs from early pipeline stages are more susceptible to contract violations. This observation is crucial because \ML~pipelines often have inherent dependencies between pipeline stages.
Violating contracts for APIs from early pipeline stages can lead to errors propagating to subsequent stages. We speculate from our data that further investigation in this area is needed. For instance, a possible future direction could be designing a verification system (as in prior work~\cite{sankaran2017darviz,dvijotham2019efficient}) with \ML~contract knowledge. Contracts could explain cases where a bug in the \ML~system is caused by an API that has a location early in the pipeline compared to where the error is registered. Catching the errors early in an \ML system can enable better performance and help reduce costs.

\finding{Training and model evaluation related APIs are the succeeding common locations that lead to contract violations across \ML~libraries.}

{\textbf{Train and Model Evaluation.}} Training is one of the stages across all libraries that are prone to contract violation. 15.2\% of the contract violations occur in APIs designed to train models. One of the primary reasons behind this is that current \ML~API documentation is insufficient on the topic of effects of optional parameters on the model's accuracy rate. Contracts can document the appropriate relationship in regard to accuracy for such optional parameters in \ML APIs found in our study. As an instance, one  \SO~post\footnote{https://stackoverflow.com/questions/24617356/} author talks about using the \scikit~API \texttt{linear\_model.SGDClassifier,partial\_fit()} due to dealing with the large size of training data. However, the \ML API user was unaware that the required contract is to shuffle the data provided as the arguments for this API. User's unawareness here can be considered towards insufficient documentation. Additionally, we observe a rate of 15.9\% in terms of model evaluation stage related contract violations. Training and model evaluation stages are significantly important since, together, they can explain the trustworthiness of the model.

\finding{Model initialization as contract violation stage is mostly prevalent in DNN libraries.}

{\textbf{Model Initialization.}} Model initialization is the stage where DNN APIs are susceptible to violating contracts. The contract violations at this stage generally show a correlation towards the crash and bad performance effects. An example of model initialization stage discussed in one \SO~post\footnote{https://stackoverflow.com/questions/47167630/}, the argument for \texttt{keras.backend.set\_session} API should be a \tf~\texttt{session}. We group this example under the model initialization stage since the API in question sets up the environment. Automated tools (e.g., Auto-Net \cite{mendoza2019towards}) that can build DNNs without human interventions can make use of API contracts from this stage to perform better and avoid crashes.
\subsection{Threats to Validity}
\label{ssec:threat}

{\textbf{Internal threats}}. The first internal threat to the validity of our results is the classification scheme we used to identify \ML~contracts. To alleviate this threat, we have prepared the classification on top of well-established contract categories \cite{10.1109/ASE.2009.60, khairunnesa2017exploiting, 10.1145/1127878.1127884}.
We have followed an open coding scheme only to add categories novel and \ML-specific. 
The group effort to create the categories helped to make consistent choices. 

To avoid the internal threat of bias in labeling attempts after training, the labelers performed an independent study, and the Kappa coefficient was used to measure inter-rater agreement. A moderator was present during the reconciliation of disagreements between raters. \newline

{\textbf{External threats}}. 
The first external threat to validity is the reliability of the dataset we have used to conduct the empirical study. There are two sources of this threat: data source and data quality.
For the data source, we have collected our data from a popular Q\&A forum for software developers, \textit{StackOverflow} (\SO). \SO, as a forum, maintains a strict moderation policy, promotes a peer-reviewing mechanism, and incorporates a reward system for encouraging quality answers from the developers\footnote{https://stackoverflow.com/help/reputation}. Moreover, the latest software developer survey on the usage of \SO forum reveals that its users come from all walks of managerial hierarchy, countries, experiences, age groups, expertise, races, etc\footnote{https://survey.stackoverflow.co/2022/}. As such, a huge user base, an abundance of topics, and a way to benchmark the quality of the contents make SO a frequent source in many SE studies~\cite{10.1145/3180155.3180260, 10.1145/3338906.3341186, Aghajani2019SoftwareDI, Beyer2014AMC, Barua2012WhatAD, Rosen2015WhatAM, Cummaudo2020InterpretingCC}. Therefore, SO represents real-world \ML developers (\ML API users) and their concerns well and makes an ideal candidate for our study. 

Next, to ensure good quality posts, we have gathered \SO~posts
that have a high enough score \cite{islam19} in terms of questions and includes an accepted answer.

We have collected \SO~posts from four top \ML~libraries; however, the number of posts that we collected varies by the library. To measure the impact of this imbalance in the dataset, we have performed a two-tail test (inequality test) on the contract types for each library.  Here, based on the \emph{t\--Stat}, and \emph{t\--Critical\--two\--tail} values, $0.178 < 3.182$, the observed difference between the sample means is small enough 
to say that the average number of contracts obtained from the four different \ML~libraries do not differ significantly. This result indicates that even though the dataset seems unbalanced in terms of the posts' frequency, the contract distribution is not unbalanced to a statistically significant extent. Additionally, prior works have~\cite{10.1145/2884781.2884800, 10.1145/3106237.3119875} recognized \SO~as an important source to extract documentation for other domains. Multiple factors have enhanced the collected \SO post used in this study compared to these prior related works. For example,  the SO posts collected were from the years 2008 and 2021 and thus contained more recent posts than those used for the paper’s submission. We also added some additional filtering criteria that suited the paper's main goal, e.g., the accepted answer post has a score higher than five and was required to contain code snippets or description that potentially describes a contract. We furthermore note that we closely followed the guidelines from prior works to conduct our study; however, there must have been some common SO posts that these previous works have studied.

Next, the nature of the methodology requires extensive manual work; thus, the number of libraries we could study is another closely related external threat. To lessen this threat, we have studied the highly-discussed four \ML~libraries based on \SO~trends since 2008. We have also observed that the number of curated posts for other \ML~libraries are less significant compared to the libraries that we have studied. For example, the libraries \emph{apache-spark-mllib} and \emph{weka} have only fourteen and two posts, respectively, for which contracts are relevant.

Furthermore, the \SO~posts are mostly about contract violations, and the answer posts talk about the needed contracts posing another external threat. If certain contracts (or violations) are not present in the dataset, our study will not find them. In essence, this is an out-of-the-vocabulary problem that is common in data mining techniques. Another possible external threat source is the need for validating findings with surveys and software developer interviews. While additional validations could raise confidence in the results, it is mitigated by the strict filtering criteria we use. We only look at the answers where at least five more users agree with the answers than those that disagree (which have been used in the past as a measure of the reliability of the answer~\cite{islam19}). Moreover, we ensure that the only accepted answers by the questioner are studied. Thus, our filtering ensures that the derived contracts reflect a consensus among the questioner, the responder (via \emph{acceptance tag}), and at least five users (minimum answer score pf 5).

Finally, we must also consider \ML API users expertise in our dataset as a threat to external validity. We have used a reliability score to mitigate this threat. Instead of using the general expertise of a software developer, the reliability score measures expertise on \ML~libraries. In short, the expertise of an user counts if they have earned the reputation from answering \ML library related questions.

\section{Related Work}
\label{related}

No previous empirical studies have investigated the requirements for \ML~API contracts, but some prior work studied related issues.

{\textbf{Studies of Bugs in \ML~Programs}}.
\cite{10.1145/3213846.3213866}  and \cite{islam19} 
have studied bugs for different DNN
libraries using two sources: \emph{Github} and \SO. They have studied frequent bugs found in DNN libraries, root causes, 
and effects of these bugs. \cite{dfaults} presented a broad taxonomy of faults that occur in \ML~systems. To that end, they have surveyed \ML~developers in addition to studying code from \emph{Github} and \SO. Their taxonomy contains a category \emph{API} that broadly categorizes usage faults of \ML APIs. However, this category is too general to apprehend different types of API contracts. 
A recent work by~\cite{islam20repairing} studied the 
challenges DNN developers face as they debug and subsequently examined the adopted bug fix patterns. \cite{10.1109/ISSRE.2012.22} performed an empirical study on general \ML libraries. 
In addition to this, the study by~\cite{jia2020tfbugs} examines the bugs found in \tf programs. 
However, all of these prior works only
present a classification for bugs; they do not identify the types of contracts that would prevent such bugs.
In contrast, we focus on the contracts that the APIs from these libraries require
and present a classification to identify different types of contracts.
Contracts differ from bug patterns in that contracts do not just 
document incorrectness; they capture conditions needed to ensure 
correct behavior. 
Contracts can also be used to assign blame: if the client violates 
the contract for an API, then the client is to blame for incorrectness/bug 
in the software. 
On the other hand, if the client satisfies its part of an API contract, 
but the API does not satisfy its part, then the API's implementation itself is buggy.

{\textbf{Classification of Contracts}}.
The notion of contracts for APIs is well-established. 
Essentially two kinds of contracts, behavioral and temporal, are most often discussed in the literature
\cite{10.1109/ASE.2009.60, 10.1145/1858996.1859035, nguyen2014mining, khairunnesa2017exploiting, 10.1145/1831708.1831723, 10.1145/1595696.1595767, 10.1145/1287624.1287632}. 
These two classes are behavioral and temporal contracts. 
In our work, we build upon these classes of contracts and 
explored their application to \ML library APIs. 
Building on an existing classification scheme helped us not to reinvent known ideas \cite{10028142446} related to API contracts. We also highlighted the new categories of specifications that are different than the non-ML APIs.

A recent study~\cite{leavens2022further} points out the lessons we have learned in the course of the JML projects. It helps to design specification languages and tools for object-oriented languages such as Java and other languages. However, this work does not provide insight into the classes of contracts that Machine learning APIs require and their similarity and dissimilarity to traditional contracts that our work focuses on.
Another research~\cite{inferICSE} proposes a technique to infer formal contracts from the natural language text of API documents. Such methodology will not suffice for \ML APIs as we illustrate that most \ML software exhibits crashes, and includes bad performance and incorrect functionality not obtained in the API documentation. Hence, we studied \SO posts and characterized the types of \ML contracts. Recently, \cite{inferISTAA} proposed a technique to extract DL-specific input constraints from API documentation and to test APIs guided by such input constraints. However, our study pointed out that there are other kinds of contracts specific to \ML, such as temporal contracts found in model architecture or other inter and intra-argument contracts, which could still be investigated further in the \ML domain.
\section{Conclusion and Future Work}

\ML~has been applied in many software systems, including critical systems. However, like non-\ML software, \ML~software can also be buggy.
The presence of bugs gives rise to the problem of improving the reliability of software that uses \ML libraries. 
\ML software can suffer degradation of reliability in a statistical sense that may not cause obvious failures, thus detecting improper use of \ML APIs can help improve its reliability.
This motivated us to perform a comprehensive study to understand the types of contracts needed for \ML~APIs.
Our study provides a taxonomy for \ML~API contracts and for violation location of these contracts. In this study, the question posts provided us with the \ML API contract violations and the accepted answer posts contained the contracts. The frequent contract violations by the \ML API users indicates the type of contracts that require immediate support. We have extracted 413 informal \ML API contracts. End-users, including people teaching the application of \ML~libraries, can directly use the informal contracts from our study, as informal API documentation. The \SO~questions indicate a need for such contracts. Additionally, language designers can use these informal contracts as examples. The extracted contracts are labeled with the taxonomy presented in this paper. To help \ML API users, libraries can be released with contracts enforced leveraging this taxonomy.

Our study has presented several key insights.
First, many required contracts for \ML~libraries are not different than traditional contracts. However, \ML API users struggle to maintain these contracts due to lack of domain knowledge, incomplete or ambiguous documentation, etc.
Second, there are distinct ML-specific contracts, e.g., ML type checking.
Additionally, \ML~APIs demonstrate a coupling between behavioral contracts and temporal contracts.
Moreover, the uniqueness of these contracts allow the client to choose either temporal ordering or a state change.
Third, \ML API users struggle with maintaining temporal method orders (especially ``eventually'' constraints) 
for \ML~APIs. 
Fourth, \ML API users often fail to satisfy input-related contracts of \ML~APIs, making input violations the most frequent root cause of contract violations in \ML~APIs.
Fifth, when the \ML~contract violations lead to system failures, the error
messages are often inadequate.
Finally, a high percentage of contract violation occurs at early ML pipeline stages. In essence, the contract violation in an \ML~API that is used in early pipeline stages may delegate the effect in subsequent pipelines. The \ML~APIs from model construction, data preprocessing, etc. can benefit more from supporting contract checking compared to \ML~APIs that are used in later pipeline stages.

From this study, we envision several future directions.  
The classification described in our study could be used to design \ML~contract specification and verification tools. Such tools could help avoid or detect API-related bugs in \ML programs or certify that an \ML~program is correct. 
An understanding of contract violations' root causes and effects described in this paper could enable better debugging mechanisms and help detect contract violations. Comprehending the difficulty of resolving certain \ML~contract violations can help in designing a recommendation system for \ML API users. For instance, a recommendation system to automatically assign difficult contract violation related questions to expert users can be designed. Finally, understanding why \ML API users make contract violations can help the designers of \ML~libraries to develop APIs that are easier to use and less prone to error.

\balance

%\section{Introduction}
%\label{intro}
%Your text comes here. Separate text sections with
%\section{Section title}
%\label{sec:1}
%Text with citations \cite{RefB} and \cite{RefJ}.
%\subsection{Subsection title}
%\label{sec:2}
%as required. Don't forget to give each section
%and subsection a unique label (see Sect.~\ref{sec:1}).
%\paragraph{Paragraph headings} Use paragraph headings as needed.
%\begin{equation}
%\end{equation}
%
%% For one-column wide figures use
%\begin{figure}
%% Use the relevant command to insert your figure file.
%% For example, with the graphicx package use
%  \includegraphics{example.eps}
%\end{figure}
%%
%% For two-column wide figures use
%\begin{figure*}
%% Use the relevant command to insert your figure file.
%% For example, with the gr63r65 aphicx package use
%  \includegraphics[width=0.75\textwidth]{example.eps}
%% figure caption is below the fi]gure
%\end{figure*}
%\begin{table}
%\begin{tabular}{lll}
%\hline\noalign{\smallskip}
%\end{tabular}
%\end{table}

\begin{acknowledgements}
This material is based upon work supported by the National Science Foundation under Grant CCF-15-18897, CNS-15-13263, CCF-17-23215, CCF-17-23432, CCF-19-34884, CNS-17-23198, CCF-22-23812. Any opinions, findings, and conclusions or recommendations expressed in this material are those of the authors and do not necessarily reflect the views of the National Science Foundation.
\end{acknowledgements}

\section*{Data Availability Statement}

The dataset (labeled \SO posts, queries, source codes, etc.) generated during the study are available in the \emph {figshare} repository, \url{https://figshare.com/s/c288c02598a417a434df}. 

\section*{Conflict of interest}

The authors declare conflict of interest with the people affiliated with Iowa State University, University of Central Florida, and Bradley University.

\bibliographystyle{spbasic}      % basic style, author-year citations
\bibliography{emse23.bib}
%\begin{thebibliography}{}
%%
%% and use \bibitem to create references. Consult the Instructions
%% for authors for reference list style.
%\end{thebibliography}

\end{document}